\documentclass[aps,pre,epsf,superscriptaddress,amsmath,amssymb,amsfonts,twocolumn,showpacs]{revtex4-1}

\usepackage{graphicx}
\usepackage{epsfig}
\usepackage{dcolumn}
\usepackage{bm}
\usepackage{braket}
\usepackage{amsmath}
\usepackage{color} 
\newcommand{\abs}[1]{\left| #1 \right|}

\usepackage{tcolorbox}
\usepackage{tabularx}
\usepackage{array}
\usepackage{colortbl}
\tcbuselibrary{skins}

\newcolumntype{Y}{>{\raggedleft\arraybackslash}X}

\tcbset{tab1/.style={fonttitle=\bfseries\large,fontupper=\normalsize\sffamily,
colback=yellow!10!white,colframe=red!75!black,colbacktitle=white!40!white,
coltitle=black,center title,freelance,frame code={
\foreach \n in {north east,north west,south east,south west}
{\path [fill=red!75!black] (interior.\n) circle (3mm); };},}}

\tcbset{tab2/.style={enhanced,fonttitle=\bfseries,fontupper=\normalsize\sffamily,
colback=yellow!6!white,colframe=red!50!black,colbacktitle=yellow!20!white,
coltitle=black,center title}}

\begin{document}
\title{Many-Body Dissipative Flow of a Confined Scalar\\ Bose-Einstein Condensate 
Driven by a Gaussian Impurity}

\author{G.C. Katsimiga}
\affiliation{Zentrum f\"{u}r Optische Quantentechnologien,
Universit\"{a}t Hamburg, Luruper Chaussee 149, 22761 Hamburg,
Germany}  

\author{S.I. Mistakidis}
\affiliation{Zentrum f\"{u}r Optische Quantentechnologien,
Universit\"{a}t Hamburg, Luruper Chaussee 149, 22761 Hamburg,
Germany}

\author{G.M. Koutentakis}
\affiliation{Zentrum f\"{u}r Optische Quantentechnologien,
Universit\"{a}t Hamburg, Luruper Chaussee 149, 22761 Hamburg,
Germany}\affiliation{The Hamburg Centre for Ultrafast Imaging,
Universit\"{a}t Hamburg, Luruper Chaussee 149, 22761 Hamburg,
Germany}

\author{P. G. Kevrekidis}
\affiliation{Department of Mathematics and Statistics, University
of Massachusetts Amherst, Amherst, MA 01003-4515, USA }

\author{P. Schmelcher}
\affiliation{Zentrum f\"{u}r Optische Quantentechnologien,
Universit\"{a}t Hamburg, Luruper Chaussee 149, 22761 Hamburg,
Germany} \affiliation{The Hamburg Centre for Ultrafast Imaging,
Universit\"{a}t Hamburg, Luruper Chaussee 149, 22761 Hamburg,
Germany}

\date{\today}

\begin{abstract}
The many-body dissipative flow induced by a mobile 
Gaussian impurity harmonically oscillating within 
a cigar-shaped Bose-Einstein condensate is investigated. 
For very {small and large driving frequencies}
the superfluid phase is preserved. 
Dissipation is identified, for intermediate driving frequencies, by the non-zero value of the drag force whose 
abrupt increase {signals the spontaneous downstream emission of an array of gray solitons.
After each emission event, typically each of the solitary waves formed decays and splits into two daughter gray solitary waves
that are found to be robust propagating in the bosonic background for large evolution times.} 
In particular, a smooth transition towards dissipation is observed, 
with the {\it critical} velocity for solitary wave formation
depending on both the characteristics 
of the obstacle, namely its driving frequency and width
as well as on the interaction strength. 
The variance of a sample of single-shot simulations indicates the fragmented 
nature of the system; here it is found to increase during evolution for driving frequencies 
where the coherent structure formation becomes significant.
Finally, we demonstrate that for fairly large particle numbers in-situ single-shot images directly capture the 
gray soliton's decay and splitting.  
\end{abstract}

\maketitle

\section{Introduction}

Dark solitons are fundamental nonlinear excitations that are found to spontaneously emerge 
in diverse physical systems ranging form nonlinear optics~\cite{zakharov,drummond,yuridavies} 
to repulsively interacting one-dimensional (1D)
Bose-Einstein condensates (BECs)~\cite{pethick,stringari,djf,siambook}
and from water waves~\cite{chabchoub} to magnetic materials~\cite{colostate}.
In the BEC setting, there exist numerous distinct
mechanisms of spontaneous generation 
of these solitonic structures that have been theoretically proposed and
also experimentally implemented.  
These density depleted states can be formed by e.g. imprinting a phase
distribution (or a density one, or both) in the 
BEC~\cite{burger,Denschlag,becker},
in interference experiments, e.g. during the collision of two condensates~\cite{weller,theo,Reinhardt,Scott}, 
or by perturbing the BEC with localized impurities moving
relative to the condensate~\cite{dutton,Engels_2007}.

In this latter context dark soliton generation induced by the motion of an impurity through the BEC
has been intensely studied~\cite{Hakim,Leboeuf,Pavloff,Brazhnyi,Radouani,Carretero,Hans,Syafwan}, 
and it can be connected with the onset of dissipation~\cite{Frisch,Winiecki,Astra}.
Landau's criterion sets the bound below which the flow remains dissipationless and no excitations are present in the 
system~\cite{Landau}. This bound for dilute BECs is the Bogoliubov speed of sound. 
However, numerous of the aforementioned theoretical and experimental studies have tested this criterion 
and estimations of significantly smaller critical velocities have been 
reported~\cite{Ketterle_1999,Ketterle_2000,Engels_2007},
being attributed to the confinement geometry and/or finite temperature effects.

In the above investigations impurities of different shape and width have been considered,
showcasing that above an obstacle dependent critical velocity gray solitons, i.e. moving dark solitons,
and sound waves
(see e.g. Refs.~\cite{Radouani,Engels_2007} and references therein) may
be generated. In fact,
under suitable conditions,
more complicated dispersive shock wave patterns may also be
formed~\cite{Hakim,Kamchatnov} 
(see also here Refs.~\cite{El1,El2,Hoefer} for higher dimensional settings
and~\cite{hoefer2} for a recent review of the latter theme of research).
This structure formation occurs whenever the velocity of the obstacle becomes locally larger than the
(local) speed of 
sound~\cite{Carretero,Frisch,Winiecki,Huepe}, and ceases to exist for high speed 
impurities~\cite{Radouani,Engels_2007,Pavloff}.
However, numerous questions still remain for which a many-body (MB) treatment of the problem has been suggested 
concerning e.g., the presence of dissipation even for very
small obstacle velocities~\cite{Astra}, or the prediction of a smoother transition towards 
dissipation~\cite{Ketterle_1999}. 
Additionally, in this latter context, the spontaneous generation of the so-called quantum 
dark solitons~\cite{sachadark3,sachadark2,sachadark1,mishmash1,mishmash2,martin,sacha,sacha17} is still an open question. 
These quantum dark solitons are known to be fragmented entities, i.e. being more adequately described by more than 
a single particle state, especially so when instabilities --at the single
particle state level-- or time-dependent
dynamical processes are involved~\cite{Katsimiga,lspp,lgspp}. 
Thus, yet another interesting prospect is to examine how the aforementioned spontaneous generation of these 
states is related to the onset of fragmentation.

Motivated by earlier experiments on this topic~\cite{Ketterle_1999,Ketterle_2000} 
but also by the ongoing interest regarding the notion of dissipation in quantum 
fluids~\cite{Wright,Ryu,Eckel,Jendrezejewski,Weimer,Roati} 
in the current effort we explore the breaking of superfluidity and the 
subsequent soliton generation upon considering 
in the MB framework the influence of an oscillating repulsive Gaussian impurity penetrating the bosonic cloud.
The specific setup examined herein has been partially motivated
by an earlier mean-field (MF) study where the controllable
dark soliton creation was demonstrated~\cite{planist}.
We provide direct numerical evidence of smaller critical velocities compared to the MF scenario for the onset 	
of dissipation and the subsequent coherent structure formation. 
In particular, it is shown that this critical velocity depends on the trapping geometry, 
the characteristics of the impurity, 
and the interaction strength and thus the corresponding speed of sound~\cite{Ketterle_2000}. 
It is found that the trap significantly lowers this 
critical value~\cite{Fedichev}, when compared to the untrapped scenario,
a feature that may be expected on the basis of its variable
density locally modifying the speed of sound. 
Moreover, wider obstacles and also effectively denser clouds 
result again in the significant decrease of this critical velocity.
More importantly, we further verify earlier suggestions of a smoother transition
towards dissipation~\cite{Ketterle_1999}  
that naturally emerges herein when MB effects are taken into account.

To systematically study the MB driving dynamics~\cite{Mistakidis_per,Mistakidis_per1} 
of a Gaussian impurity penetrating a 1D 
harmonically confined bosonic cloud, we 
use the Multi-Configuration Time-Dependent Hartree Method for bosons 
(MCTDHB)~\cite{cederbaum1,cederbaum2} and contrast our findings with MF theory. 
We investigate the system's dynamical response covering the range from weak to strong driving frequencies. 
To infer about the dissipative flow we inspect the drag force~\cite{Astra,khamis,Kato,Pinsker} exerted 
on the fluid by the Gaussian impurity. 
For very small or large driving frequencies the superfluid is dissipationless. 
We demonstrate that dissipation, occurring for intermediate driving frequencies, is followed by the 
spontaneous downstream emission, with respect to the impurity's motion, of an array of moving gray 
solitary waves that naturally emerge in a MB environment and
{dispersive sound waves moving upstream.}
This important outcome reveals, further adding to the existing studies regarding the aforementioned solitary 
waves~\cite{sachadark3,sachadark2,sachadark1,mishmash1,mishmash2,martin,sacha,sacha17}, 
that such nonlinear excitations can be dynamically excited in a MB system and are thus of fundamental origin.
Each of these solitary waves soon after its formation is found to decay~\cite{sachadark3,sachadark2,Katsimiga,lspp} 
and split into two daughter gray solitary waves that remain robust propagating in the BEC background
for large evolution times.
Additionally, utilizing single-shot simulations, the fragmented nature of the system is
showcased probed by the evolution of the variance of a sample of single-shots. 
In particular, fragmentation is found to be maximal when dissipation occurs, a result 
that holds as such for a wide parametric window.
{Furthermore, we provide direct 
numerical evidence of the aforementioned gray solitons' generation and more importantly their 
subsequent decay and splitting in the present type of a dynamical MB setting.}
We showcase the latter by simulating in-situ single 
shot images upon considering a fairly large bosonic cloud. 
Finally, we retrieve superfluidity for high speed impurities~\cite{Paris}  
namely for velocities being almost six times the bulk (i.e. the maximal value measured around the trap center) speed of 
sound.

The presentation of our work is structured as follows. In Section \ref{setup} the model setup and 
the MB ansatz are provided. 
Section \ref{dynamics} contains our numerical findings both in the single orbital MF case
and in the MB correlated approach. 
In Section \ref{single_shots} we offer experimentally testable
evidence of the observed MB evolution by simulating 
single-shot images. 
Our results are summarized in Section \ref{conclusions}, along with interesting directions for future study. 
Finally, in Appendix A we discuss the convergence behavior of our numerical results.

\section{Driving Scheme and many-body ansatz }\label{setup}

In the following we consider the MB quantum dynamics of a scalar harmonically confined 1D BEC being relaxed to
its ground state, with a Gaussian impurity located initially ($t=0$) 
at the right edge of the cloud ($x>0$) where the density is almost vanishing,
and subsequently crossing the BEC towards its left edge ($x<0$). 
Such a cigar-shaped geometry is experimentally realizable upon considering a strong confinement 
along the perpendicular $y, z-$directions, namely $\omega_{\perp}>>\omega_x\equiv\Omega$. 
To study the out-of-equilibrium dynamics of this system we consider the following 
form for the driving potential
\begin{eqnarray}
V_{D}(x,t)= V_{ext}(x) + V_G(x,t).
\label{driving}
\end{eqnarray}
In Eq.~(\ref{driving}) the first term is the standard parabolic potential, $V_{ext}(x)=\left(1/2\right) m \Omega^2 x^2$, 
of strength $\Omega$, while $m$ is the particle mass.
The second part corresponds to the mobile Gaussian impurity $V_G(x,t)=\frac{A}{\sqrt{2 \pi} w}
\exp\left\{-\frac{[x-x_D(t)]^2}{2 w^2}\right\}$. Such time-dependent localized
potentials may stem from an intensely focused laser beam spot moving through the
condensate~\cite{Ketterle_1999,Ketterle_2000}. Here, $A>0$ is the amplitude of the Gaussian
impurity corresponding to a repulsive potential for the atoms and $w$ is its width.
Initially, the system is in its ground state for $x_D(t=0)=B$. The initial position
of the obstacle, $B\approx R_{TF}$, is located at the right edge of the bosonic cloud
possessing approximately a Thomas-Fermi profile of the form 
$\rho^{(1)}(x;0)\approx \sqrt{m\Omega^2\left(R^2_{TF}-x^2\right)/2g_{1D}}$.
The Thomas-Fermi radius reads $R_{TF}=\sqrt{\frac{2 \mu}{m \Omega^2}}$ (see Fig.~\ref{Fig:1}), where
$\mu$ is the chemical potential, while $g_{1D}$ is the effective interaction strength (see below).  
At $t=0$ the impurity commences an oscillatory motion $x_D(t)=B
\cos(\omega_D t)$ with frequency $\omega_D=2\pi/T_D$, for either half ($0 \le t \le t_1=T_D/2$) or
one ($0 \le t \le t_1=T_D$) oscillation period corresponding to a single or double crossing of the
impurity through the BEC respectively. For $t>t_1$ the impurity
remains stationary at $x_D(t)= B
\cos(\omega_D t_1)$ and the system is left to evolve in the absence of external driving. Note here
that the velocity of the obstacle during $0 \le t \le t_1$ reads $u_D=-B\omega_D
\sin(\omega_D t)=\mp \omega_D \sqrt{B^2 -x_D^2(t)}$. It then follows that the ratio of $u_D$ and the
unperturbed local speed of sound at the position of the impurity is 
$c_0\big(x_D(t)\big)=\sqrt{g m^{-1} \rho^{(1)}\big(x_D(t);0\big)}$, 
and it is approximatively constant throughout the driving. 
Thus, we can relate the values of $\omega_D$ with the characteristic (and commonly employed) velocity ratio
$u_D(x_D)/c_0(x_D)\equiv u_D/c_0 \approx \sqrt{2} \omega_D/ \Omega$, since $B\approx R_{TF}$. 

The MB Hamiltonian consisting of $N$ bosons each with mass $m$ trapped in the 1D potential of Eq.~(\ref{driving})   
reads 
\begin{equation}
\begin{split}
H(x_1,x_2,...,x_N;t)= & \sum_{i=1}^{N} \left[ -\frac{\hbar^2}{2 m} \partial_{x_i}^2  
+V_{D} \left(x_i;t \right) \right]\\ 
&+g_{1D}\sum_{i<j} \delta(x_i - x_j).\\
\end{split}
\label{Eq:SDsoli0}
\end{equation}
Operating in the ultracold regime, the short-range delta  
interaction potential between particles located at 
positions $x_i$ can be adequately described by $s$-wave scattering. 
The effective interaction strength in this case is defined as 
$g_{1D}=\left(2\hbar^2 a_0/ma^2_{\perp}\right)\left(1- \frac{|\zeta(1/2)|a_0}{\sqrt{2}a_{\perp}} \right)^{-1}$~\cite{g1d},
where $a_{\perp}=\sqrt{\hbar/\left(m \omega_{\perp}\right)}$ 
is the transverse harmonic oscillator length characterized by 
$\omega_{\perp}$, and $a_0$ denotes the free space 3D $s$-wave scattering length.
In the present investigation we consider the dynamics of repulsively interacting bosons namely $g_{1D}>0$. 
Experimentally $g_{1D}$ can be adjusted either via $a_0$ utilizing  
magnetic or optical Feshbach resonances~\cite{Inouye,Chin} or through the corresponding $\omega_{\perp}$
utilizing confinement-induced resonances~\cite{g1d}. 
Our setup can be to a good approximation realized by considering e.g. a gas of $^{87}$Rb atoms.
To render the Hamiltonian of Eq.~(\ref{Eq:SDsoli0}) dimensionless the following transformations are performed for the 
energy, length and time scales: $\tilde{H}= \left(\hbar \omega_{\perp}\right)^{-1}~H$, 
$\tilde{x}= \sqrt{\left(m \omega_{\perp}\right)/ \hbar}~x$, and 
$\tilde{t}= \omega_{\perp}~t$. According to the above the interaction strength, Gaussian amplitude and width
are expressed in dimensionless units of $\tilde{g}= \sqrt{m/ \left(\hbar^3 \omega_{\perp}\right)}~g_{1D}$,
$\tilde{A}= \sqrt{m/ \left(\hbar^3 \omega_{\perp}\right)}~A$, and $\tilde{w}= \sqrt{\left(m \omega_{\perp}\right)/ \hbar}~w$ 
respectively. 
For convenience henceforth we shall omit the tildes, 
and all the numerical values given are to be understood in the 
aforementioned dimensionless units.
Furthermore, throughout this work the trapping
frequency, $\Omega=0.1$, the Gaussian amplitude, $A=1/2$ and the width, $w=0.1$ (being always smaller
than the corresponding healing length $\xi=1/ \sqrt{2 \mu}$), are held fixed unless it is stated
otherwise. 
We remark here that larger values of either $A$ or $w$ lead to a decrease of the critical velocity 
and an enhanced number of emitted solitons~\cite{Hakim}.
Therefore, we are left with two free parameters namely $g$ and $\omega_D$. 
\begin{figure}[ht]
\includegraphics[width=0.43\textwidth]{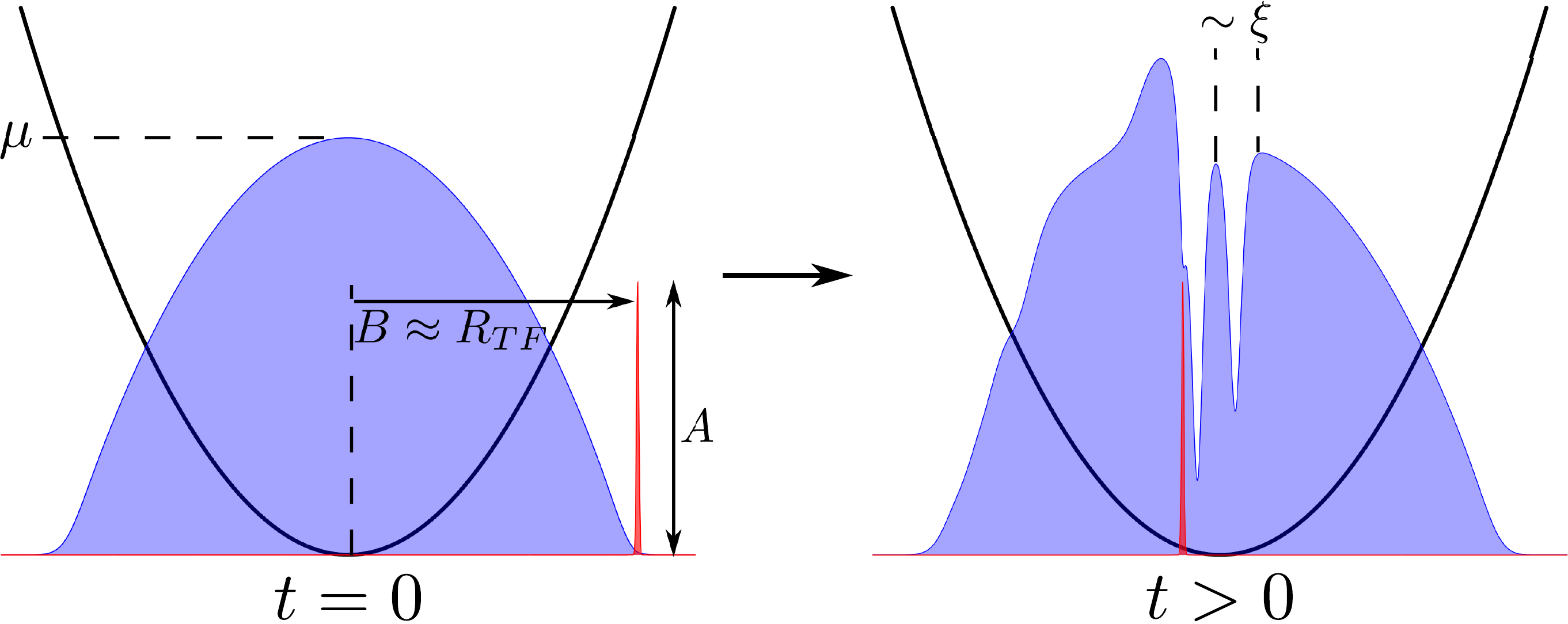}
\caption{(Color online) Schematic representation of the driving dynamics, 
which leads to the spontaneous generation of two, in the sketch shown, downstream dark solitons.
Notice that the width of the impurity
(denoted by the thin red Gaussian curve)
is much smaller than the corresponding healing length. } \label{Fig:1}
\end{figure}

To systematically take into account the important particle correlations inherent to the system 
we utilize the MCTDHB approach \cite{cederbaum1,cederbaum2}, which is a reduction of the more general 
Multi-Layer-Multi-Configuration Hartree method for bosonic and fermionic Mixtures (ML-MCTDHX)~\cite{Cao,matakias,moulosx}. 
The MB wavefunction of the system $\Psi_{MB}(x_1,\dots, x_N;t)$, with ($x_1,\dots,x_N$) labelling the spatial coordinates 
of the atoms, is constructed by permanents that are built upon $M$ distinct time-dependent single particle functions (SPFs)
\begin{equation}
\begin{split}
&\Psi_{MB} (x_1,\dots, x_N;t)= \sum_{\substack{n_1,\dots,n_M\\ \sum n_i=N}} A_{(n_1,\dots,n_M)}(t)\times \\ 
&\sum_{i=1}^{N!} \mathcal{P}_i
 \left[ \prod_{j=1}^{n_1} \varphi_1(x_j;t) \cdots \prod_{j=1}^{n_M} \varphi_M(x_{K(M)+j};t) \right]. \label{Eq:MCansatz}
 \end{split}
\end{equation} 
In the above expression $\mathcal{P}$ is the permutation operator exchanging the particle 
positions $x_i$, $i=1,\dots,N$, $\varphi_l(x;t)$, $l=1,2,...,M$, denote each SPF, 
$K(M)\equiv\sum^{M-1}_{k=1}n_k$,
and $A_{(n_1,\dots,n_{M})}(t)$ correspond to the time-dependent expansion coefficients of a particular permanent.   
$N$ refers to the total particle number and $n_k$ is the occupation number of the $k$-th SPF. 
Following e.g. the McLachlan time-dependent variational 
principle~\cite{McLachlan} for the generalized ansatz of Eq.~(\ref{Eq:MCansatz}) yields the MCTDHB 
\cite{cederbaum1,cederbaum2,matakias,moulosx}  
equations of motion.  
These consist of a set of $\frac{(N+M-1)!}{N!(M-1)!}$ 
linear differential equations for the expansion coefficients and $M$ nonlinear integro-differential 
equations for the SPFs $\varphi_i(x;t)$.  

The spectral representation of the one-body reduced density matrix reads 
\begin{equation}
\label{eq:4} \rho^{(1)} (x,x';t) =N \sum\limits_{i=1}^{M}
{{n_{i}}(t){\phi _{i}}(x,t)} \phi _{i}^*(x',t), 
\end{equation}
where $M$ refers to the number of natural orbitals, $\phi_i(x;t)$, used. The latter 
are the eigenfunctions of the one-body reduced density matrix~\cite{Titulaer,Naraschewski,Sakmann}
being normalized to unity, while
$n_i$ are the corresponding eigenvalues or natural populations.
Note here that the one-body density $\rho^{(1)} (x;t)\equiv \rho^{(1)} (x,x'=x;t)$, and 
for $M\rightarrow \infty$, $\rho^{(1)} (x;t)$ becomes the exact one-body density 
$\tilde{\rho}^{(1)} (x;t)$.  
Our MB wavefunction $\Psi_{MB} (x_1,\dots, x_N;t)$ reduces to the
MF one, $\Psi_{MB} (x_1,\dots, x_N;t)\rightarrow\Psi_{MF} (x_1,\dots, x_N;t)$, 
when the corresponding natural occupations obey $n_1(t)=1$, $n_{i\neq1}(t)=0$.    
In the latter case the first natural orbital $\phi_1(x;t)$ reduces to the MF wavefunction which for the 
$N$-particle system is $\Psi_{MF} (x_1,\dots,x_N;t) = \prod_{i=1}^{N} \phi_1(x_i;t)$ and obeys the Gross-Pitaevskii 
equation~\cite{stringari}. 
Finally, we remark that the above-mentioned natural populations $n_{i}(t) \in [0,1]$ characterize 
the degree of the system's fragmentation or interparticle correlations \cite{Penrose,Mueller}. 
For only one macroscopically occupied orbital the system is said to be 
condensed, otherwise it is fragmented. 

\section{Non-Equilibrium Driven Quantum Dynamics}\label{dynamics} 
\begin{figure*}[ht]
\includegraphics[width=0.78\textwidth]{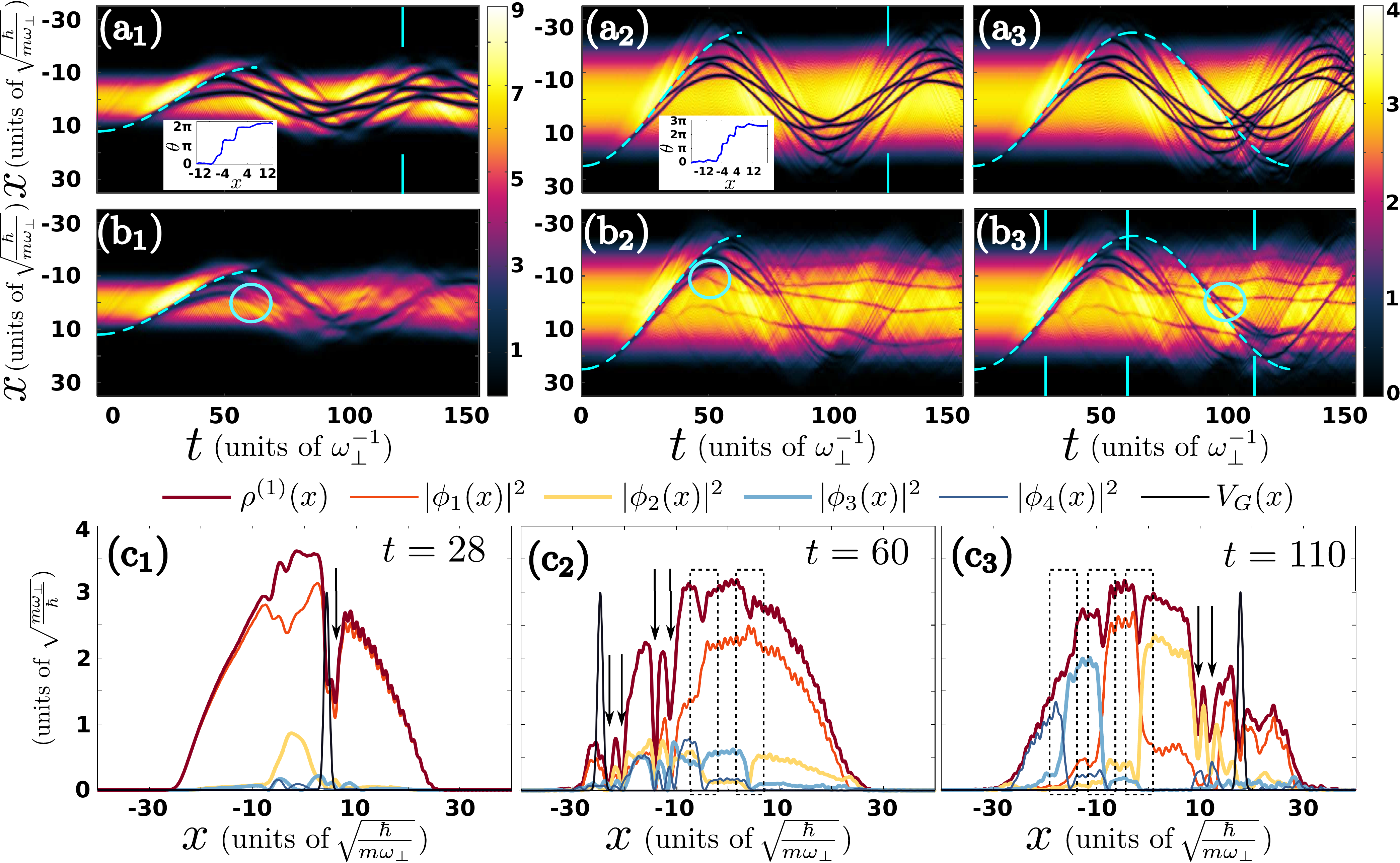}
\caption{(Color online) $(a_1)$-$(a_{2})$ [$(b_1)$-$(b_{2})$] Evolution of the one-body density, $\rho^{(1)}(x;t)$,
within the MF [MB] approach for a single crossing of the obstacle through the BEC. From left to right in each row the 
intraspecies interaction coefficient increases from $g=0.1$ to $g=1$.
$(a_3)$ [$(b_3)$] Same as the above but for an oscillation period of the obstacle. In all cases dashed curves indicate the 
position of the Gaussian impurity and circles showcase the decay events observed in the MB scenario.  
Insets illustrate the phase profile, $\theta(x)$,
for the MF case, at $t=120$ [see the vertical cyan lines in $(a_1)$, $(a_2)$] 
after all the solitons are generated.
Notice that there is a series of jumps, none of
which is exactly $\pi$ because none of the produced MF gray solitons is stationary. 
$(c_1)$- $(c_3)$ Profile snapshots of $\rho^{(1)}(x;t)$, as well as of the densities $\abs{\phi_i(x,t)}^2$  
with $i=1,\dots,4$, 
of the four natural orbitals used in the MB approach for $g=1$ for an oscillation period at different time instants of the 
evolution indicated by the solid cyan lines in $(b_3)$. 
Arrows here, indicate the location of the gray solitons while dashed rectangles
capture the formation of the domain-walls building between the higher-lying orbitals. 
The Gaussian impurity amplitude and width in these cases are magnified to provide better visibility.   
The initial ground state configuration contains $N=100$ bosons and the trapping frequency is fixed to $\Omega=0.1$.
$B=25$ ($B=12$) for $g=1$ ($g=0.1$) so as to initialize the Gaussian of width $w=0.1$ and amplitude $A=0.5$ at the right edge 
of the TF cloud.}
\label{Fig:2}
\end{figure*}

A direct comparison of the MF and the MB driven  
dynamics can be deduced by inspecting the spatio-temporal evolution of the one-body density, $\rho^{(1)}(x;t)$, 
shown in Figs.~\ref{Fig:2} $(a_1)$-$(a_3)$
and $(b_1)$-$(b_3)$ respectively.
In all cases the obstacle is initially placed at
$x\gtrsim R_{TF}$, the oscillation frequency is $\omega_D=0.05$,
while $B=25$ ($B=12$) for $g=1$ (for $g=0.1$).

In both approaches, it is observed that from the very early stages of the out-of-equilibrium dynamics 
a density imbalance between the left and the right side of the impurity during its motion occurs.
This density imbalance results in the spontaneous emission 
of both a downstream (behind the impurity) and an upstream (in front of the impurity)
disturbance with respect to the obstacle's motion. 
This emission is activated when the velocity of the impurity becomes greater  
than the local speed of sound ($c(x)=\sqrt{g\rho^{(1)}(x)}$).
Note that despite placing the impurity at extremely low 
densities our used profile of the velocity ensures that at the beginning of the driving dynamics the impurity is subsonic, 
i.e. $u_D<c$.  
The disturbance behind the driver arises in the form of gray solitons, 
while the one in front of it consists of dispersive sound waves.
In particular, three and four almost regularly 
spaced solitons are clearly emitted by the impurity for a single crossing
(i.e., $t_1\in$ [0, $T_D/2$]), 
of the obstacle through the condensate in the MF case illustrated in Figs.~\ref{Fig:2} $(a_1)$ and $(a_2)$ 
for $g=0.1$ and $g=1$ respectively (see also Table~\ref{table}). 
Each of the solitary waves formed develops a characteristic phase jump
{always smaller than $\pi$ 
as it is evident in the corresponding  
phase profiles, $\theta (x)$, depicted as insets in 
Figs.~\ref{Fig:2} $(a_1)$ and $(a_2)$ for $g=0.1$ and $g=1$ respectively at $t=120$.} 
The number of {solitary waves generated} in both approaches is presented in Table~\ref{table} 
for different driving frequencies and upon increasing the interaction strength.
Notice that in both approaches for fixed $\omega_D$  the number of
{coherent structures} 
emitted by the source increases for increasing $g$.
We remark here that among the excitations formed, 
we exclude all waves having a numerically 
identified speed $u_{sol}\gtrsim 0.95c$, since for these states we cannot assign a clear phase jump and thus
cannot definitively distinguish them from sound waves.

In the corresponding MB dynamics
{the following key differences} when compared to the MF scenario are discernible. 
The emission of solitary waves remains the same as in the MF approximation for small $g$ values but is slightly 
larger upon increasing $g$ with the solitons emitted being five in the MB approach instead of 
four in the MF case (see again here Table~\ref{table}, for $\omega_D=0.05$
and $g=1$). 
Furthermore, and also independently of the magnitude of the interaction strength most of the 
solitons soon after their formation decay in the MB 
scenario~\cite{sachadark3,sachadark2,sachadark1,sacha,sacha17,lgspp,lspp}. 
This decay is followed by a splitting of each gray solitonic structure into two daughter solitary waves.
Case examples of such decay and splitting events are indicated with solid circles in Figs.~\ref{Fig:2} $(b_1)$-$(b_3)$.
These daughter states are more robust and propagate in the BEC background for large evolution times.
The aforementioned process repeats itself when considering the dynamics for a full period of oscillation, 
with the spontaneous emission of downstream gray solitons and upstream dispersive sound waves taking place
whenever the obstacle's motion becomes locally supersonic, {i.e. $u_D>c$ 
[see Figs.~\ref{Fig:2} $(a_3)$, $(b_3)$].} 
\begin{table}\centering
\begin{tabular}{|c||c|c|c|c|}
 \hline
 \multicolumn{5}{|c|}{{\bf Solitary Wave Counting}} \\
 \hline \hline 
$\omega_D$ &~~~0.02~~~ & ~~~ $0.05$~~~ & ~~~ $0.08$~~~ &~~~  $0.2$~~~ \\
 \hline \hline
 MF $g$=1.0  & $0$ & 4 &  5 &  0\\
 MB $g$=1.0 & 0 & 5 &  6 &  0\\
 MF $g$=0.1  & 0 & 3 &  2 &  0\\
 MB $g$=0.1  & 0 & 3 & 2 & 0 \\
 \hline 
\end{tabular}
\caption{Solitary wave counting for a single crossing of the impurity through the BEC, upon increasing both the driving frequency 
and the interparticle interaction strength. Note that among the excitations formed we exclude all waves having a numerically 
identified speed $u_{sol}\gtrsim 0.95c$. For such high speeds no clear phase jump can be attributed to these structures as can 
be deduced by inspecting the insets in Figs.~\ref{Fig:2} $(a_1)$ and $(a_2)$.}
\label{table}
\end{table}
To gain further insight into the generation of gray solitons 
within the MB approach in Figs.~\ref{Fig:2} $(c_1)$-$(c_3)$ profile snapshots of the one-body density 
as well as for the four natural orbitals used are presented for initial, intermediate, and longer 
evolution times for $g=1$. 
{In the one-body density presented in all Figs.~\ref{Fig:2} $(c_1)$-$(c_3)$, 
some of the density dips correspond to solitary wave structures associated predominantly with
the first orbital. See e.g. the gray soliton generated in $\rho^{(1)}(x;t=28)$ indicated by a black arrow in  
Fig.~\ref{Fig:2} $(c_1)$, that is clearly supported by a density dip developed in the first orbital.
Others, especially so at later times, and most
notably so at $t=110$ depicted in Fig.~\ref{Fig:2} $(c_3)$ may be associated with higher orbitals such as
the second one (see again black arrows here). 
However, a key observation emerging from the breakdown
of the one-body density through the orbitals is that {\it not all}
substantial density dips in the MB case are associated with gray
solitons, contrary to what would be the case in the MF scenario.
Instead, numerous ones among them, notably at later times arise
due to domain wall structures~\cite{siambook,Trippenbach} 
not only of the first with the second orbital, but also of the second with the third and so on. 
These domain walls are indicated by dashed rectangles in Figs.~\ref{Fig:2} $(c_2)$ and $(c_3)$.}
\begin{figure*}[ht]
\includegraphics[width=0.76\textwidth]{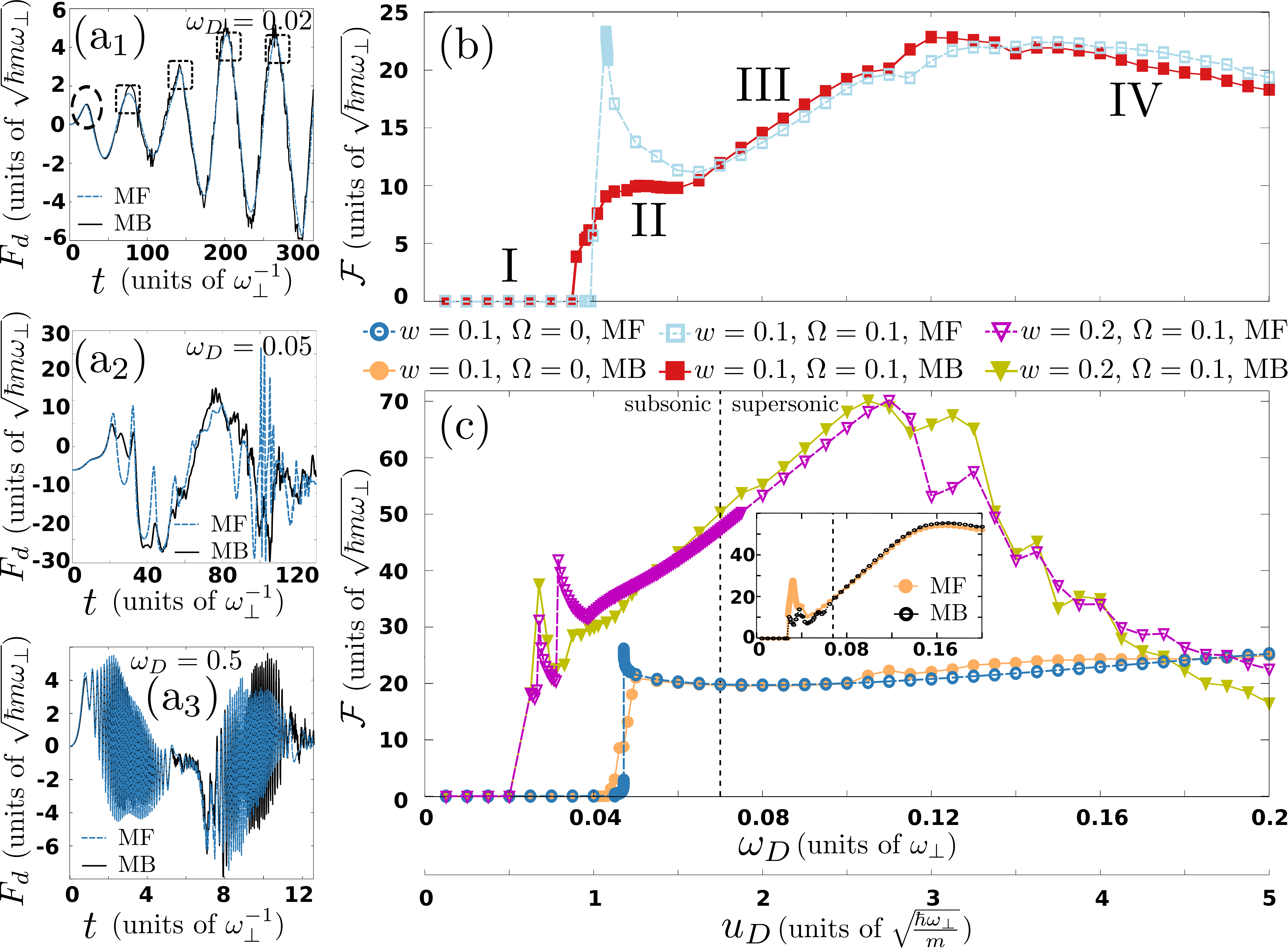}
\caption{(Color online) $(a_1)$-$(a_3)$ Evolution of $F_d(t)$ for a full oscillation period for small, 
intermediate, and large driving frequencies $\omega_D$ respectively in both the MF and the MB case. 
Note that for small and large $\omega_D$ the dynamics is 
well-described by MF theory, while alterations between the two approaches occur at intermediate frequencies. 
The dashed circle and squares indicate the injection peak and the oscillation maxima respectively.
$(b)$ Maximum drag force $\mathcal{F}$ as a function of the driving frequency $\omega_D$ in both approaches (see 
legend). 
The dynamics is displayed in four regions, namely I-IV (see text).
$(c)$ The same as $(b)$ but for different parametric variations and both within the MF and the MB approach (see legend).
In all cases $g=1$ and $N=100$, and the second axis used shows the corresponding velocities. Notice here, that the 
corresponding bulk speed of sound is $c_0=1.73$, and thus for $\omega_D>0.07$ 
the impurity is supersonic (vertical dashed line). 
The inset is the same as $(c)$ but for $g=0.1$ and $B=12$. 
$A=0.5$ ($A=1$) for $w=0.1$ ($w=0.2$).} \label{Fig:3}
\end{figure*}

{\subsection{Characterizing the Dissipative Flow via the Drag Force}} \label{drag1}
In order to characterize the superfluid or dissipative flow past the obstacle 
we invoke the drag force exerted on the fluid by the moving impurity.
The drag force~\cite{Pavloff,khamis}
{in our inhomogeneous setting is defined as} 
\begin{eqnarray}
F_d(t)=\int^{+\infty}_{-\infty} dx \rho^{(1)}(x;t) \frac{d{V_{D}}}{dx},
\label{drag}
\end{eqnarray}
where finite drag, $F_d\neq0$, implies dissipation while $F_d=0$ corresponds to a superfluid flow. 
In particular, Figs.~\ref{Fig:3} $(a_1)$-$(a_3)$ illustrate $F_d(t)$ during an oscillation period, 
for small, intermediate, and large driving frequencies respectively. 
We remark here that from the very early stages of the driving dynamics the drag force acquires a finite value.
This is a natural by-product of the moving impurity, as the
antisymmetric nature around the impurity center of $\frac{dV_D}{dx}$
necessitates a symmetric $\rho^{(1)}(x;t)$ around the impurity in order
to vanish. 
Additionally, and also independently of the value of $\omega_D$, 
a local maximum of $F_d(t)$ at $t\approx25$ is observed 
in all Figs.~\ref{Fig:3} $(a_1)$-$(a_3)$. 
The existence of this peak, indicated by a dashed circle in Fig.~\ref{Fig:3} $(a_1)$, 
stems from the fact that the impurity penetrates the BEC from the edge of the Thomas-Fermi radius 
inducing a density imbalance imprinted in the finite value of the drag force, that would otherwise 
be absent (e.g. by initializing the impurity from the trap center).
For small driving frequencies, e.g. for $\omega_D=0.02$ depicted in Fig.~\ref{Fig:3} $(a_1)$,  
$F_d(t)$ for $t>25$ undergoes oscillations of increasing amplitude,
the maxima of which are indicated by dashed rectangles. 
This oscillating behavior is directly related with the collective dipole  
oscillation of the trapped BEC due to the presence of the impurity~\cite{Kohn,Albert},
and can be removed by subtracting the background density, and further neglecting the contribution of the trap from the 
definition of Eq.~(\ref{drag}), namely 
$\tilde{F}_d=\int^{+\infty}_{-\infty} dx\left[\rho^{(1)}(x;t)-\rho^{(1)}(x;0)\right]\frac{d{V_{G}}}{dx}$. 
However, this collective motion of the atoms is not related to the onset of dissipation in the system.
This result can be directly inferred upon inspecting the corresponding one-body density evolution 
for small driving frequencies (results not shown here for brevity) 
which reveals that solitary wave formation is absent  
during the obstacle's motion even for larger propagation times.
As such in the following we will ignore both the {\it injection peak} 
[namely the first peak in $F_d(t)$] as well as the collective oscillation 
experienced by $F_d(t)$ in our calculations, since in the present 
setup their inclusion is not connected with the onset of dissipation.

Increasing $\omega_D$ entails even more rapid oscillations of the corresponding drag force 
as can be deduced by comparing e.g. Figs.~\ref{Fig:3} $(a_2)$, and $(a_3)$. 
Notice however, the significantly lower values of $F_d(t)$ for large driving frequencies e.g. for $\omega_D=0.5$,
indicating, as we will trace later on, that for very fast oscillations of the impurity superfluidity is again 
recovered~\cite{khamis}.
In line with these significantly lower values of the drag force  
it is found that MF theory accurately describes the dynamics both for small (well below unity) 
and large $\omega_D$'s.
This result is also directly related and captured by the negligible degree of 
fragmentation present in the system when considering its MB evolution inside 
these two parametric windows (see also below).

Differences between the two approaches become significant for intermediate frequencies (or velocities), with the drag 
force acquiring rather large values reaching a global maximum of the order of $F_d(t \approx 80) \approx 18$ 
($F_d(t \approx 100) \approx 25$) in the MB (MF) approach 
presented in Fig.~\ref{Fig:3} $(a_2)$. It is for
this intermediate region that soliton formation takes place,
and the critical frequency (or velocity) of its occurrence will be estimated in what follows. 
The emission of these structures can be identified by the sharp increase of $F_d(t)$ and its subsequent decrease
as the emitted solitary waves detach from the spatial extent of the impurity.

To shed further  light on the above-discussed distinct dynamical regions, and more importantly to evaluate the
critical velocity above which the onset of dissipation occurs, we next  
consider a single crossing of the Gaussian impurity through the BEC. 
For times $t$ within the interval $t_1\in [0, T_D/2]$ 
we evaluate the maximum drag force, $\mathcal{F}\equiv{\rm max} \{F_d(t)\}$, exerted on the fluid for varying driving 
frequencies/velocities and also 
for different values of the interparticle repulsion $g$ and width $w$ of the impurity. 
Figs.~\ref{Fig:3} $(b)$ and $(c)$ summarize our findings. 
In all cases illustrated in Figs.~\ref{Fig:3} $(b)$ and $(c)$
and also in both the MF and the MB approach the following general remarks can be made. 
Four different regions can be identified corresponding roughly to small (I), intermediate (II-III), and 
large (IV) driving frequencies. 
For small driving frequencies a plateau of zero drag force is observed--recall that we neglect in this 
calculation both the injection peak as well as the collective oscillatory motion of the drag force. 
This plateau is followed by an increase and a subsequent decrease, being more pronounced in the MF approach, 
of $\mathcal{F}$ for increasing $\omega_D$ till almost the end 
of the subsonic regime ($\omega_D\approx0.07$).  
Note that the bulk speed of sound, i.e. its maximal value calculated around the center of the trap,
is $c_0(0)\approx1.73$ e.g. for $g=1$, and thus for frequencies $\omega_D>0.07$ 
the impurity is supersonic.
This decreasing tendency is followed by an almost linear increase of $\mathcal{F}$ that reaches its maximum
value within the supersonic regime [see the dashed vertical lines 
in Fig.~\ref{Fig:3} $(c)$] dropping down to almost $20\%$ of this maximum
for even larger driving frequencies. 
Overall we can conclude that 
the small $\omega_D$ effect is associated
with the motion being too slow, essentially adiabatic and hence solitary waves cannot be generated. 
On the other hand very large driving frequencies lead to 
an ``averaging out'' effect where the oscillation is too fast to excite coherent
structures.

\begin{figure}[ht]
\includegraphics[width=0.47\textwidth]{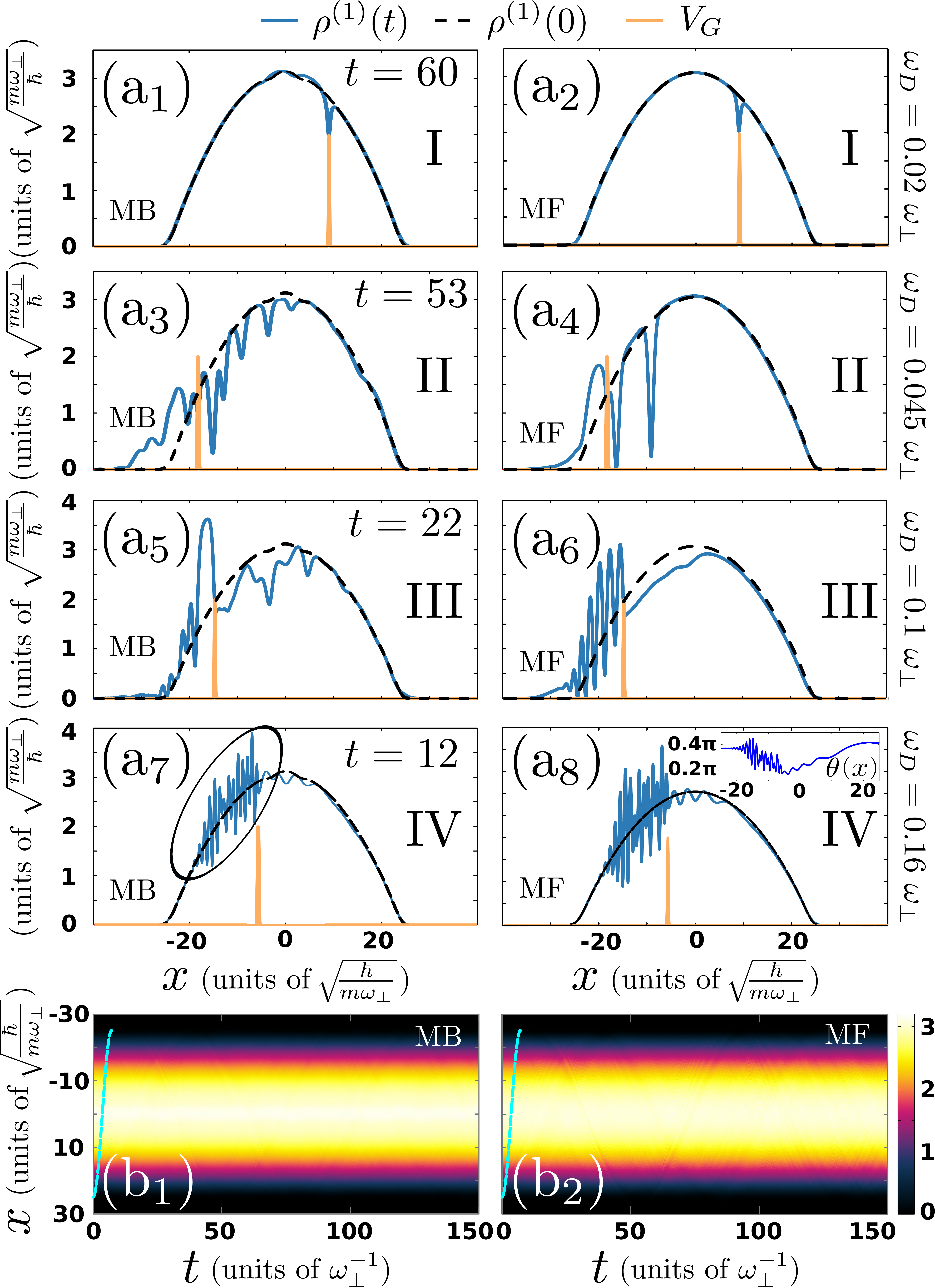}
\caption{(Color online) 
($a_i$) Profile snapshots of the one-body density at selected time instants during evolution in all four, I-IV, 
dynamical regions identified in Fig.~\ref{Fig:3} both within the MB and the MF approach, with $i=1,\dots,8$ being odd 
for the former and even for the latter case respectively (see legends). 
The inset in ($a_8$) within the MF approximation,
indicates the corresponding phase, 
$\theta(x)$, to showcase the rapid oscillations of the dispersive sound wave formed. 
($b_1$) and ($b_2$) illustrate the spatiotemporal evolution of the one-body density in the MB and the MF approach 
respectively for $\omega_D=0.5$ when superfluidity is restored. 
In all cases $g=1$ and $N=100$.} 
\label{Fig:3a}
\end{figure}

Profile snapshots of both the MB and the MF driving dynamics in each of the above-mentioned regions
are illustrated in Figs.~\ref{Fig:3a} $(a_1)$-$(a_8)$ for $g=1$.
For small driving frequencies (region I) depicted in Figs.~\ref{Fig:3a} $(a_1)$ and $(a_2)$ for the MB and the MF case 
respectively, 
the flow remains dissipationless during the impurity's motion
until a {\it critical} frequency (region II) is reached. 
The latter is found to be $\omega_{D_c}=0.04$
corresponding to a critical velocity $u_{c}/c_0 \approx 0.57$ within the MF
approximation a result that is in excellent agreement with earlier theoretical predictions~\cite{Hakim},   
while it is estimated to be {$\omega_{D_c}=0.036$} (with $u_{c}/c_0 \approx 0.52$)
in the MB scenario. 
Within this region (II) the flow becomes dissipative
imprinted as a sharp jump in $\mathcal{F}$ estimated in the MF case,
in contrast to the smoother transition observed in the MB scenario.
In particular focusing in the vicinity of the critical point we deduce that dissipation is enhanced in
the MB when compared to the MF case. This result is indicated by the higher growth rate of
$\mathcal{F}$ in the former approach [see Fig.~\ref{Fig:3} $(c)$].
Inspecting the MF evolution at $u_D \approx u_c$ 
it is found that the solitons are emitted in the low-density region (periphery)
of the bosonic cloud~\cite{Carretero}.
For increasing $\omega_D$, $\mathcal{F}$ exhibits a sharp peak at $\omega_D=0.043$,
indicating that solitons are emitted also around the trap center. 
It is this peak that is absent in the MB case which instead showcases a smoother transition (in line with earlier
experimental predictions~\cite{Ketterle_1999}). This observation further suggests that the 
critical velocity, estimated to be slightly smaller in the MB case~\cite{Ketterle_2000}, depends more 
strongly on the density in the MF case rather than in the MB scenario.
The dissipative flow in this region is accompanied by the aforementioned 
almost periodic emission of downstream gray solitons and upstream dispersive sound waves
in both approaches [see e.g. Figs.~\ref{Fig:3a} $(a_3)$ and $(a_4)$]. 
After each soliton emission for a given frequency a decrease in the drag force
occurs (which is of course lifted once the soliton travels sufficiently
far from the defect, hence the repeated emission observed), 
while upon further increasing the driving frequency but remaining in region II, 
leads to the emission of a higher number of coherent structures. 

Further increasing the driving frequency, such that the driving becomes supersonic, we enter region III. Here, 
a coexistence of solitary and dispersive sound waves
is observed with the number of the former decreasing gradually while the amplitude of the latter is enhanced 
[see Figs.~\ref{Fig:3a} $(a_5)$ and $(a_6)$ for the MB and the MF outcome respectively]. 
In both approaches, gray solitons cease to exist for $\omega_D\gtrsim 0.12$, namely within region IV where $\mathcal{F}$ 
reaches also its maximum value, and the dynamics is dominated
by significantly amplified sound waves being repeatedly emitted by the obstacle. 
These sound waves are indicated by the ellipse in Fig.~\ref{Fig:3a} $(a_7)$.
For comparison we also show in this region the corresponding MF profile in Fig.~\ref{Fig:3a} $(a_8)$.
Notice that in this case within this train moving upstream we can distinguish a front and a rapidly oscillatory tail. 
Inspecting the corresponding phase, $\theta(x)$, shown as an inset in Fig.~\ref{Fig:3a} $(a_8)$ we observe that it also 
exhibits an oscillatory behavior. 
Additionally here, small phase jumps are still imprinted in the phase being connected with the downstream motion of 
excitations present in the fluid. However these phase jumps correspond to velocities $u_{sol}\gtrsim 0.95c$. 
It is the presence of these waves that leads to the finite, though much smaller, value of the drag 
force for the half period driving considered herein, indicating that even for high 
impurity speeds we do not yet recover superfluidity. 
The latter is retrieved for even larger driving frequencies than those 
depicted in Fig.~\ref{Fig:3} $(c)$ estimated to be $\omega_D\gtrsim0.25$. A case example of 
the observed dynamics for these large driving frequencies 
is illustrated in Figs.~\ref{Fig:3a} $(b_1)$ and $(b_2)$ 
for the MB and the MF approach respectively. Indeed in this case we observe
that excitations are absent both in the MB and MF cases.

The above distinct dynamical regions (I-IV) are shifted towards smaller 
or larger $\omega_D$ values depending on the characteristics of the
obstacle, the spatial inhomogeneity induced by the considered trapping geometry, 
and on the interaction strength (and thus the speed of sound~\cite{Ketterle_2000}).  
In particular, and as depicted in Fig.~\ref{Fig:3} $(c)$
larger widths of the Gaussian impurity significantly reduce the critical frequency (velocity) for coherent structure 
formation e.g. for $w=0.2$ we find that
$\omega_{D_c} \approx 0.025$ ($u_{c}/c_0 \approx 0.36$). The oscillatory behavior of $\mathcal{F}$ in this case 
at intermediate driving frequencies and also in both approaches is related to the fact that 
the number of solitary waves generated 
increases dramatically when compared to the $w=0.1$ case.
Additionally, the critical velocity is increased in the corresponding
homogeneous setting~\cite{Ketterle_2000}, while for smaller interaction strengths and thus 
smaller speed of sound again a decrease in the critical velocity is observed 
[see the inset in Fig.~\ref{Fig:3} $(c)$]. 
In particular, in the homogeneous setting the obstacle oscillates in a region of uniform density, resulting to the 
observed differences in the overall shape of $\mathcal{F}$ in both approaches when compared to the 
trapped scenario [see Figs.~\ref{Fig:3} $(b)$ and $(c)$]. 
In this way, the critical frequency for soliton formation is 
$\omega_{D_c} \approx 0.044$ ($u_{c}/c_0 \approx 0.64$) in the MB case,
and $\omega_{D_c} \approx 0.047$ ($u_{c}/c_0 \approx 0.68$) in the MF approach, leading to a wider region I  
when compared to the confined case, while
all II-IV regions are shifted to even larger $\omega_D$ values~\cite{comment}. 
Note here that for $g=0.1$ deviations between the MF and the MB approaches are negligible for almost every $\omega_D$ 
but in the region of significant
{solitary wave} formation, see e.g. $\omega_D\approx0.03$. 
\begin{figure*}[ht]
\includegraphics[width=0.8\textwidth]{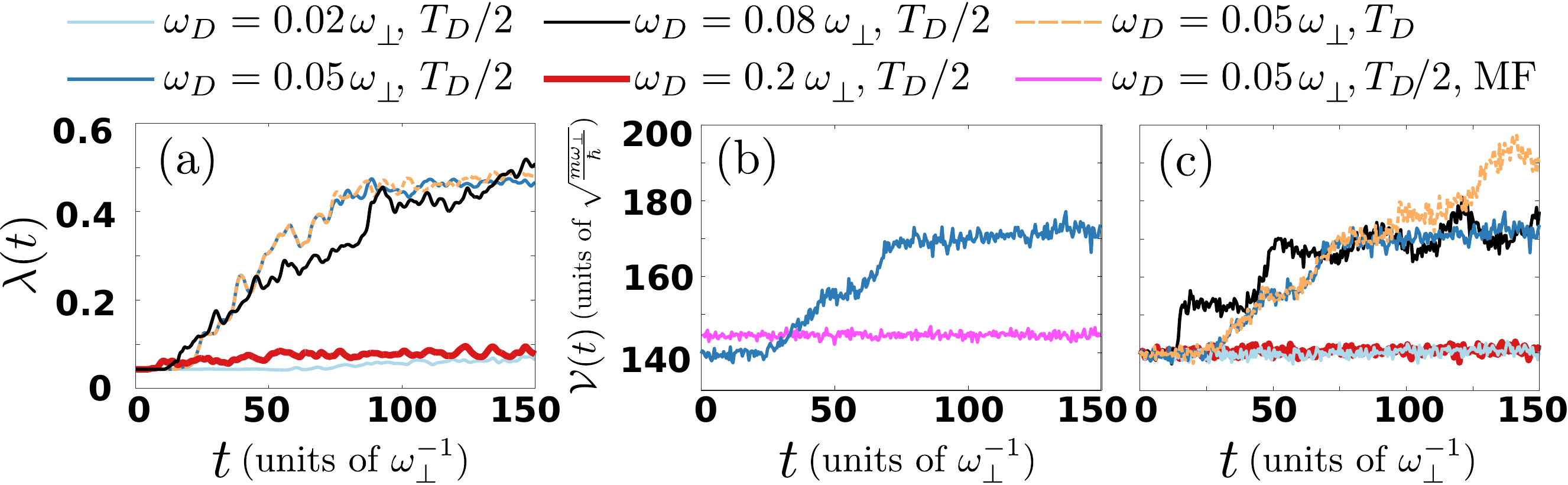}
\caption{(Color online) 
$(a)$  Deviation from unity of the first natural population during evolution for different driving 
frequencies and also for a half and a full oscillation period of the obstacle (see legend). $(b)$ Evolution of the variance 
$\mathcal{V}(t)$ of a sample of single shots, $N_{shots}=500$, in both {the MF and the MB approach} (see legend). 
$(c)$ $\mathcal{V}(t)$ in the MB scenario and for all the cases illustrated in $(a)$.
Other parameters used are the same as in Fig~\ref{Fig:2}.} \label{Fig:4}
\end{figure*}

To expose the degree of correlations \cite{Mistakidis_per,Mistakidis_per1,negative,spatial} 
inherent in the system's evolution,
and corresponding departure from the MF regime
for the different driving frequency ranges
discussed above, 
we next rely on the deviation from unity of the first natural population, $\lambda(t)=1-n_1(t)$. 
Fig.~\ref{Fig:4} ($a$) shows $\lambda(t)$ for different $\omega_D$'s and $g=1$.  
Recall that a state with $n_1 (t) = 1$ ($\lambda(t)=0$) is referred to as fully coherent or condensed, while 
if $n_i(t)$ [$\lambda(t)$] significantly deviates from unity [zero] the more modes $n_i(t)\neq0$ ($i=2,3,4$ here) 
are populated resulting to a fragmented state \cite{Penrose,Mueller}. 
In our setup $\lambda(0)\approx0.03\neq0$ holds implying that the initial (ground) state is already depleted. 
Let us first focus on the driving protocol that refers to half of an oscillation period of the obstacle through the 
condensate. 
As time evolves, fragmentation is generally present being more pronounced at moderate driving frequencies 
residing in regions II and III (see e.g. $\omega_D=0.05$) where the
structure formation is significant. 
Contrary to that, for either small (see e.g. $\omega_D=0.02$) 
or large ($\omega_D=0.2$) driving frequencies 
residing respectively in regions I and IV, for which all excitations are absent 
or the dynamics is dominated by dispersive sound waves, 
$\lambda(t)$ is significantly suppressed. 
In all cases the maximum fragmentation rate, 
identified by the slope $\delta \lambda(t)\equiv\left(\lambda(t+\Delta t)-\lambda(t)\right)/\Delta t$, 
occurs during the driving (i.e. $t<T_D/2$) while after the single crossing of the impurity through the BEC 
$\lambda(t)$ changes in a far less dramatic manner. 
Finally, and upon considering the driving dynamics for a complete period of oscillation of the impurity,
we observe that $\lambda(t)$ is exactly the same as that resulting from a single obstacle crossing for $0<t \leqslant T_D/2$. 
However, for $t>T_D/2$ slight deviations between the free dynamical evolution of the system and the driven one occur.
\vspace{0.4cm}

\section{Single-shot simulations}\label{single_shots} 
Having discussed the degree of the system's fragmentation we next showcase how the correlated character of the driven 
dynamics can be inferred by performing in-situ single-shot absorption measurements~\cite{lspp,zorzetos,Lode} 
which essentially probe the spatial configuration of the atoms being dictated by the MB probability distribution. 
Relying on the MB wavefunction being available within MCTDHB we emulate the corresponding experimental 
procedure and simulate such in-situ single-shot images at each instant of the evolution. 
This simulation procedure is well-established (for more details e.g. see \cite{lgspp,Lode,kaspar,Chatterjee}) 
and therefore it is only briefly outlined below. 
Referring to the time, $t=t_{im}$, of the imaging we first calculate $\rho^{(1)}(x;t_{im})$ 
from the MB wavefunction $\ket{\Psi_{N}}\equiv \ket{\Psi_{N}(t_{im})}$. 
Then, a random position $x'_1$ is drawn obeying $\rho^{(1)}_N(x'_1;t_{im})>l_1$ where $l_1$ is    
a random number in the interval [$0$, $\max\{\rho^{(1)}_N(x;t_{im})\}$]. 
Next, one particle located at a position $x'_1$ is annihilated and the $\rho^{(1)}_{N-1}(x;t_{im})$  
is calculated from $\ket{\Psi_{N-1}}$. 
To proceed, a new random position $x'_2$ is drawn from $\rho^{(1)}_{N-1}(x;t_{im})$. 
Following this procedure for $N-1$ steps we obtain the distribution of positions 
($x'_1$, $x'_2$,\dots,$x'_{N-1}$) which is then convoluted with a point spread function  
resulting in a single-shot image $\mathcal{A}(\tilde{\textbf{x}})$, where $\tilde{\textbf{x}}$ denote the 
spatial coordinates within the image. 
We note that the employed point spread function, being related to the experimental resolution,  
consists of a Gaussian possessing a width $w=1 \ll l\equiv\sqrt{1/\Omega}=10$. 
$l$ denotes the harmonic oscillator length. 
\begin{figure}[ht]
\includegraphics[width=0.47\textwidth]{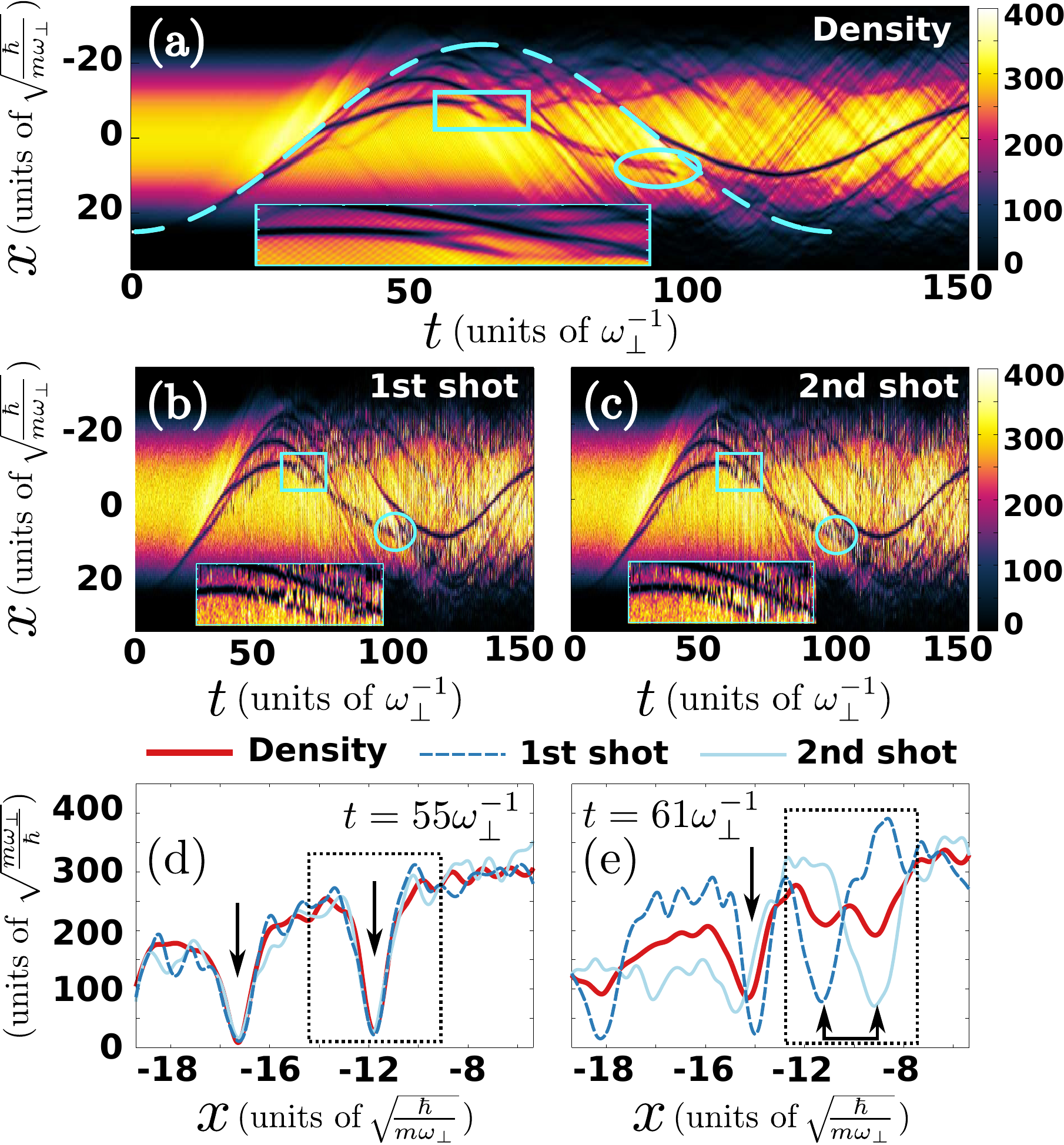}
\caption{(Color online) 
$(a)$ Spatio-temporal evolution of the one-body density for $N=10^4$ bosons.  
$(b)$-$(c)$ Evolution of the first and the second in-situ single-shot images respectively.
Solid {squares and circles} indicate the decay events occurring during propagation that are 
evidently captured by the in-situ single shot images depicted in $(b)$-$(c)$. 
To better visualize a decay event the insets in ($a$), ($b$), ($c$) depict the spatiotemporal evolution within the region 
marked by the corresponding square. Profiles of the one-body density and of both shots $(d)$ prior and $(e)$ right after the 
first decay event marked with squares in $(a)$-$(c)$. Dashed rectangles here indicate the region of interest (see text), 
while black arrows show the location of the gray solitons formed.
Other parameters used are the same as in Fig~\ref{Fig:2}.} \label{Fig:5}
\end{figure}

To estimate the role of fragmentation from single-shot measurements we utilize their variance 
for each time instant of the driven dynamics. 
Below, and unless stated otherwise, we mainly refer to the dynamics 
induced by the mobile impurity upon considering its single crossing  
through the condensate, namely for times up to $t_1=T_D/2$.  
The variance of a sample of $N_{shots}$ single-shot measurements $\{\mathcal{A}_k(\tilde{\textbf{x}})\}_{k=1}^{N_{shots}}$ reads 
\begin{equation}
\mathcal{V}(t_{im})=\int d\tilde{\textbf{x}} \frac{1}{N_{shots}} \sum_{k=1}^{N_{shots}} 
[\mathcal{A}_k(\tilde{\textbf{x}};t_{im})-\bar{\mathcal{A}}(\tilde{\textbf{x}};t_{im})]^2,  
\end{equation}
where $\bar{\mathcal{A}}(\tilde{\textbf{x}};t_{im})=(1/N_{shots}) \sum_{k=1}^{N_{shots}} 
\mathcal{A}_k(\tilde{\textbf{x}};t_{im})$. 
$\mathcal{V}(t)$ over $N_{shots}=500$ is illustrated in Fig. \ref{Fig:4} ($b$) both at the MF and the MB level for $\omega_D=0.05$.  
As it can be deduced in the MF approximation $\mathcal{V}(t)$ remains almost constant exhibiting small amplitude oscillations 
during evolution. 
This can be attributed to the fact that for a product MF ansatz all particle detections 
are independent from each other and the corresponding single-shots of such a state merely reproduce the 
one-body density (see also the discussion below). 
In contrast, when correlations are included an overall 
increase of $\mathcal{V}(t)$ is observed.
An increase that is more pronounced during the obstacle's crossing ($t<T_D/2\approx63$), resembling this way the 
fragmentation rate $\delta \lambda(t)$ as it can seen by comparing Figs. \ref{Fig:4} ($a$) and ($b$). 
This similarity between the fragmentation process and $\mathcal{V}(t)$ has already been observed in several 
MB investigations~\cite{Lode,lgspp,lspp} and can be explained as follows. 
Referring to a perfect condensate, i.e. $n_1(t)=1$, $\mathcal{V}(t)$ is almost constant during the evolution 
since all the atoms in the corresponding single-shot measurement are picked from the same SPF $\varphi(t)$ 
[see the discussion below Eq.~(4)]. 
Contrary to that, for a MB correlated system the corresponding MB state is a superposition  
of several mutually orthonormal SPFs $\varphi_i(t)$, $i=1,...,4$ [see Eq. (\ref{Eq:MCansatz})]. 
A superposition changes drastically the variance of a sample of single-shots dissociating it from 
its MF counterpart, since in this case the 
distribution of the atoms in the cloud depends strongly on the position of the already imaged ones.
Moreover, the fact that $\mathcal{V}(t)$ increases during evolution is attributed also to the build-up of higher-order 
superpositions in the course of the dynamics. 
To expose further the relation between the growth rate of $\mathcal{V}(t)$ and the fragmentation rate $\delta \lambda(t)$, we 
present in Fig. \ref{Fig:4} ($c$) $\mathcal{V}(t)$ calculated solely in the MB approach for different driving 
frequencies $\omega_D$. 
The impurity, here, oscillates over $T_D/2$ through the ensemble which is then left to evolve. 
It is observed that $\mathcal{V}(t)$ {resembles} the increasing tendency of $\lambda(t)$ during evolution for 
all $\omega_D$ regions as can be deduced by comparing Figs. \ref{Fig:4} ($a$) and ($c$). 
For instance, $\mathcal{V}(t)$ is more pronounced for moderate driving frequencies (e.g. $\omega_D=0.05$) 
for which the
{coherent structure formation is most significant}, and becomes lesser in magnitude 
for either small or large $\omega_D$'s (see e.g. $\omega_D=0.02$ and $\omega_D=0.2$ respectively). 
Finally, for an impurity oscillating over one period inside the BEC $\mathcal{V}(t)$ grows further for $t>T_D/2$ when compared 
to the situation of a half period of oscillation. 

Let us now investigate whether the gray soliton generation, and its subsequent decay and splitting can be observed 
in an in-situ single-shot image. 
We remark that a direct observation of the one-body density in a single-shot image is not a-priori possible 
due to the small particle number, $N=100$, of the considered bosonic gas and the presence 
of multiple orbitals in the system. 
Within our treatment the MB state is constructed as a superposition of multiple orbitals [see Eq. (\ref{Eq:MCansatz})] 
and therefore imaging an atom alters the MB state of the remaining atoms and hence their one-body density. 
The latter is in direct contrast to a MF state, composed from a single macroscopic orbital, 
where the imaging of an atom does not affect the distribution of the rest 
(see also the above discussion of the corresponding variance). 
Thus in order to fairly capture the spatio-temporal evolution of the one-body density 
via a single-shot image, we consider the driving dynamics for the oscillating impurity over one period   
for $\omega_D=0.05$ and upon considering a fairly large number of particles, i.e. $N=10^4$. 
The one-body density evolution of this system is shown in Fig. \ref{Fig:5} ($a$) where the creation of solitary waves which 
are prone to decay is observed, as in the case of $N=100$ bosons. 
{Note, however, that the products of a decay event are much less discernible when compared to the $N=100$ particle case 
[see Fig. \ref{Fig:2} ($b_3$) and Fig. \ref{Fig:5} $(a)$] as here many collision events between the emitted solitons occur which 
distort the MB evolution. Additionally, the domain wall structures developed between the higher-lying orbitals, though still 
present, are less apparent, i.e. having significantly lower population, and as such are not clearly imprinted
in the one-body density. 
In this system $g=0.01$ which essentially corresponds to the scaled interaction strength (such that $Ng=const.$) 
of the $N=100$ with $g=1$ case. 
Also we note here that the simulation of this system has been performed within 
a two orbital approximation as the inclusion of further orbitals is computationally prohibitive. 
Figs. \ref{Fig:5} ($b$), ($c$) illustrate the first and the second
samples of simulated in-situ single-shot images $\mathcal{A}
(\tilde{\textbf{x}})$ for the entire evolution time. 
It is evident that in both shots the soliton formation takes place, 
and most importantly their subsequent decay and splitting is observed
resembling this way the overall behavior of the one-body density.
Case examples of these latter events at different time instants are indicated in Figs. \ref{Fig:5} ($b$), ($c$) 
by light blue squares and circles. 
For a better visualization of a decay event imprinted in a single-shot image during evolution we provide 
as an inset the spatiotemporal region indicated by the square. 
For this magnified region in Figs.~\ref{Fig:5} ($d$), ($e$) 
profiles of the one-body density as well as of both shots are illustrated prior (at $t=55$) 
and after (at time $t=61$) the decay and splitting respectively. 
Dashed rectangles mark the spatial position of one parent soliton [Fig.~\ref{Fig:5} ($d$)] 
which subsequently decays and splits into  
two daughter ones [Figs.~\ref{Fig:5} ($e$)] while arrows indicate the locations 
of all solitons. 
Notice that before the decay takes place both shots clearly 
capture the solitons imprinted in the one-body density 
[see Figs.~\ref{Fig:5} ($d$)]. Remarkably enough, at later times when the two  
fragments following the decay of the initial solitary 
wave are formed, the first shot ``picks up" the fragment travelling to the left (with respect to the trap center) while the 
second shot clearly captures the fragment travelling to the right [see also the inset in Fig.~\ref{Fig:5} $(a)$]. 
Importantly here, in both shots the solitons appear much more depleted 
when compared to the fragments imprinted in the corresponding one-body density which is a clean manifestation of the quantum 
dispersion of dark soliton's position that has already been reported in~\cite{sacha,sachadark1}. 
Recalling here that the fragments formed are multi-orbital 
entities this ``pick up" selection observed in the single-shots 
further implies the presence of domain walls building upon the higher-lying orbitals. 
Thus, we observe that the single-shot images not 
only fairly capture the structures building upon the one-body density [in particular compare 
the insets of Figs. \ref{Fig:5} ($a$) and ($b$), ($c$)], but via comparing consecutive shots also 
signatures of the domain wall structures present can be inferred.
Finally, notice that in the corresponding single-shot images the BEC background for 
evolution times $t>60$ becomes significantly excited 
and the emergent solitonic structures are hardly discernible. 
This noise source stems from the increasing shot-to-shot variations due to the presence of fragmentation. 
Recall that in a correlated system each shot alters the MB state \cite{Syrwid_shots}.

\section{Conclusions \& Future Work}\label{conclusions}
In the present work the MB 
dissipative flow of a harmonically confined scalar Bose-Einstein condensate 
has been investigated, exploring similarities and differences
of the latter from the single orbital MF case. 
To quantify dissipation the drag force exerted on the fluid upon driving an oscillating Gaussian 
impurity through the bosonic cloud is used as a measure. Distinct dynamical regions are 
identified corresponding to small, intermediate,
and large oscillation frequencies. 
It is found that for slow (adiabatic) and rapid (averaged
out) oscillations of the impurity, 
MF theory adequately describes the driving dynamics, a result that is clearly 
captured also by the negligible fragmentation measured in these
parameter regions. 

However, at moderate driving frequencies (or velocities)
fragmentation becomes significant and thus a MB treatment is required.
In this region an increase in the maximum of the drag force signals the
onset of dissipation.
{The critical frequency for this transition is found to be slightly smaller and the 
transition itself smoother when MB effects are taken into account}.
{In particular, the critical velocity for the onset of 
dissipation and the subsequent {solitary wave} formation is found to 
depend on the interaction strength and thus on the corresponding speed of 
sound~\cite{Ketterle_1999,Ketterle_2000}.} 
It also depends on the trapping geometry, shifting to smaller critical values the breaking of 
superfluidity when compared to the unconfined case, 
and finally on the characteristics of the impurity. 
In this latter case, deviations between the two approaches 
are much more pronounced as the interparticle interaction
is increased, with the critical velocity measured in the MB scenario 
being shifted to slightly smaller values.   
Once this critical frequency is reached a spontaneous emission of downstream gray solitons 
and upstream dispersive sound waves takes place.
This emission occurs whenever the impurity's motion becomes locally supersonic
increasing the number of solitary waves generated in an oscillation period.
We demonstrate that these states naturally emerge in the MB
setting as fundamental excitations of the system. 
Each of the gray solitary waves formed soon after its generation is found to decay and split into 
two daughter gray solitary waves that are seen to propagate for large evolution times.
Moreover, importantly at later times we identify domain wall states
that are unique to the MB phenomenology as they arise between
different orbitals. These are reflected into density dips appearing at the
one-body density, at
first glance, as gray solitons, although a closer inspection of the
different orbital densities reveals their domain wall structure.
To probe fragmentation we compare its growth rate in this region of moderate frequencies 
to the growth rate of the variance of 
a sample of single-shot simulations that are used to complement our findings. 
Importantly here, upon enlarging the number of particles present in the system the evolution of in-situ single shot images
directly dictates not only the generation, but for the first time the decay and splitting of these solitary waves offering 
evidence that can be experimentally tested. 
However contrary to the fewer particle scenario, upon enlarging the system the previously robust domain wall 
structures are no longer clearly imprinted in the one-body density evolution but their presence can be indirectly inferred
by comparing consecutive single-shot images.
The smearing effect of the domain walls in the one-body density stems from the fact that in this
case the coherent structures that are generated after the splitting, suffer multiple collisions with one another as well as 
with the sound waves present, which significantly excites the 
background rendering their observation far less straightforward.

Further increasing the driving frequency, there exists a parameter window where both solitons and dispersive sound waves
coexist, while for even larger driving frequencies the dynamics is dominated by dispersive sound waves. 
Finally, it is found that superfluidity is retrieved upon considering very high speed impurities with velocities 
estimated to be more than six times the bulk speed of sound in the MB approach. 

There is a line of interesting directions worth pursuing in future efforts. 
A straightforward one would be to generalize the current findings in two dimensions where 
the role of dark solitons is played by vortices. 
In this case, driving a Gaussian surface through the BEC may  
give rise to stripe dark states or even transient states such as oblique dark solitons~\cite{el},
whose spontaneous formation and dynamical evolution in a MB environment is yet
rather unexplored. Notice that in addition now the {\it spatial width}
of the driving impurity along the transverse direction will play a role
in the emergence (and the resulting nature) of the coherent structures
in this system.
This latter exploration seems to be particularly timely and relevant,
given the use of the motion of laser beams (in the so-called chopsticks method)
both theoretically at the MF level~\cite{bettina}, as well
as experimentally~\cite{kali}, in order
to produce arbitrary vortex configurations
in 2D. A related interconnected experimental exploration concerns
the also very recently reported formation of the two-dimensional
localized Jones-Roberts solitons~\cite{kaib}. 
Examining whether these structures
can be created through such a process at the multi-orbital
level and how their MB analogues
would dynamically evolve renders this an especially worthwhile direction
of near-future work.
Additionally, one could also consider the presence of dipolar interactions~\cite{Bland} in the 1D 
setting and further study the spontaneous generation of nonlinear excitations in such a case, and  
the subsequent deviations from the dynamics observed herein.  
Finally, it would be a challenging future task to study how the presented results are altered at finite 
temperatures~\cite{burger,Proukakis} and 
explore how fragmentation competes with thermal effects in MB systems.
Since experiments have started exploring more systematically the role
of thermally-induced dissipation towards the expulsion of coherent
structures such as vortices~\cite{shinn}, addressing such questions also for states
such as dark solitons acquires particular timeliness and significance
as a topic for future study.

\appendix
\begin{figure}[ht]
\includegraphics[width=0.45\textwidth]{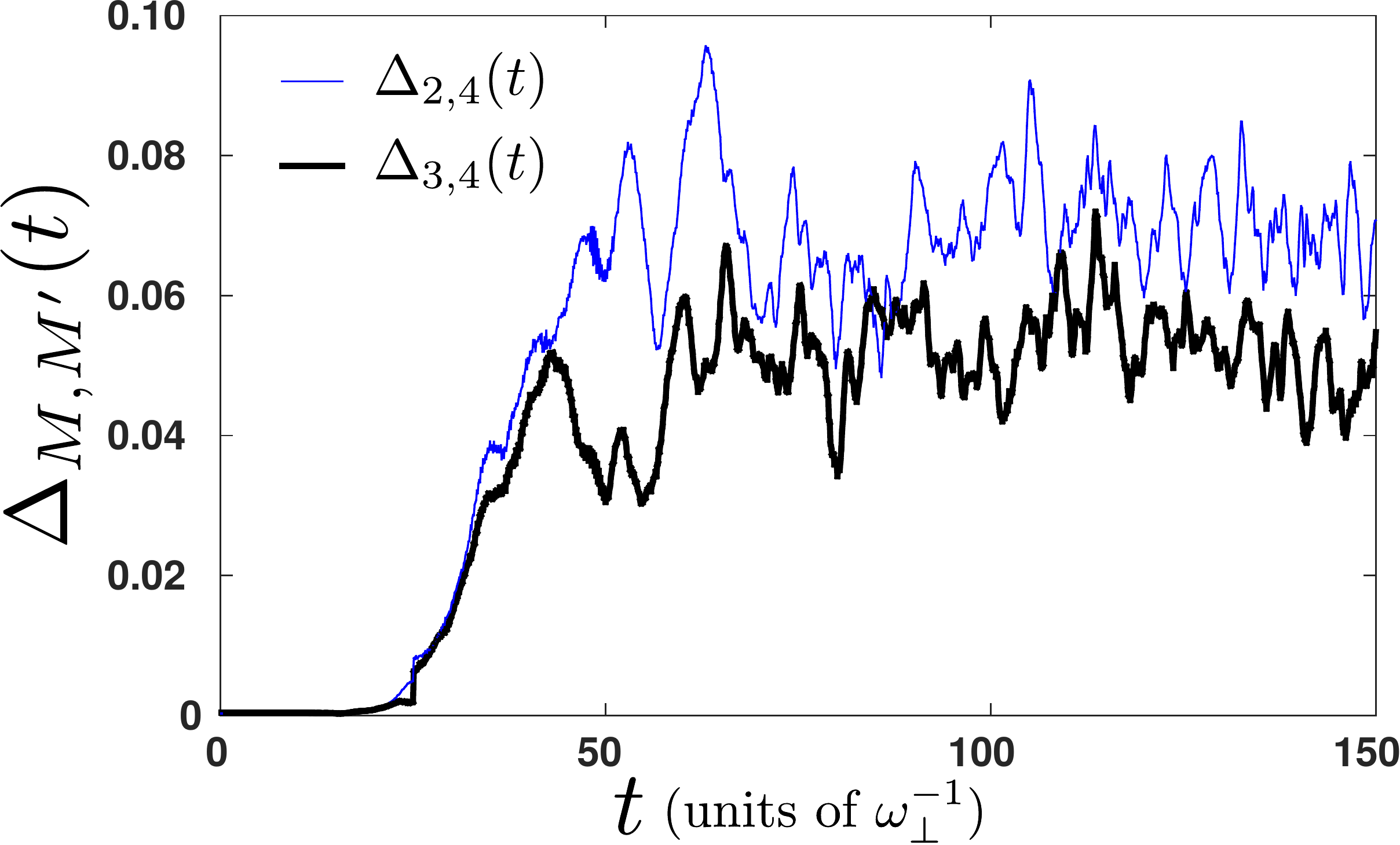}
\caption{(Color online) $\Delta_{M,M'}(t)$ during evolution upon increasing the number of orbitals used (see legend). 
$\omega_D=0.05$ and the driving dynamics corresponds to a full oscillation period of the impurity through the BEC.} 
\label{Fig:convergence}
\end{figure}

\section{Ingredients and Covergence of the Many-Body Simulations} \label{sec:numerics1}

In the present Appendix we outline some basic features of our computational method MCTDHB~\cite{cederbaum1,cederbaum2}, discuss the ingredients 
of our numerical calculations and showcase the convergence of our results. 
Let us remark that within our implementation we use the Multi-Layer Multi-Configuration 
Time-Dependent Hartree method for bosonic and fermionic Mixtures (ML-MCTDHX) \cite{Cao,matakias,moulosx}. 
It is an extended version of the MCTDHB and has been designed for the treatment of 
multicomponent ultracold systems \cite{Katsimiga,lspp,BF}.  
We note that for single bosonic species, as is the case considered herein, ML-MCTDHX reduces to MCTDHB. 

MCTDHB is a variational method for solving the time-dependent MB Schr{\"o}dinger equation of 
interacting bosonic systems. 
It relies on expanding the total MB wavefunction with respect to a time-dependent and 
variationally optimized basis, which enables us to capture the important correlation effects 
using a computationally feasible basis size. 
Namely, it allows us to span more efficiently the relevant, for the system under consideration, subspace 
of the Hilbert space at each time instant with a reduced number of basis states when compared to expansions relying on 
a time-independent basis. 
In particular, the MB wavefunction of $N$ bosons is expressed by a linear combination 
of time-dependent permanents $\ket{\vec{n}}=\left| {{n_1},{n_2},...,{n_M};t}\right\rangle$ with 
time-dependent expansion coefficients $A_{\vec{n}}(t)$.  
These permanents build upon $M$ time-dependent single-particle functions $\ket{\varphi_i(t)}$, $i=1,\dots,M$ which are 
expanded within a time-independent 
primitive basis $\{\ket{k}\}$ of dimension $\mathcal{M}$. 
We note here that for $M=1$ the MB wavefunction is given by a single permanent 
$\ket{n_{1}=N;t}$ and the method reduces to the time-dependent Gross-Pitaevskii 
MF approximation. 

For our simulations, we use a primitive basis consisting of a sine discrete variable 
representation with $\mathcal{M}=1500$ grid points. 
To perform the simulations into a finite spatial region, we impose hard-wall boundary conditions 
at the positions $x_{\pm}=\pm80$. 
The Thomas-Fermi radius of the bosonic cloud is of the order of $R_{TF}\sim25$ for $g=1$ and $R_{TF}\sim12$ for $g=0.1$. 
The location of the imposed boundary conditions does not affect our results as we never observe 
appreciable densities beyond $x=\pm 35$. 
To achieve numerical convergence, we ensure that the expectation value of the observables of interest become 
to a certain degree insensitive when increasing the number of basis states. 
Regarding our simulations we have used $M=4$ orbitals. 
To quantify the degree of convergence, for instance, of the one-body density evolution 
we invoke the spatially integrated difference between the $M$ and $M'$ orbital configurations 
\begin{equation}
 \Delta_{M',M}(t)=\frac{\int_{x_{-}}^{x_{+}} dx|\rho_{M'}^{(1)}(x,t)-\rho_{M}^{(1)}(x,t)|}{\int_{x_{-}}^{x_{+}} 
 dx\rho_{M}^{(1)}(x,t)}.  
\end{equation} 
Fig. \ref{Fig:convergence} shows both $\Delta_{2,4}(t)$ and $\Delta_{3,4}(t)$ during the evolution 
upon considering the driving of the impurity through the BEC for an oscillation period, 
for $\omega_D=0.05$ lying within region II where fragmentation becomes significant.
A systematic convergence of $\Delta_{M,M'}(t)$ is showcased for increasing the number of orbitals. 
As it is evident $\Delta_{3,4}(t)$ testifies negligible deviations between the two orbital configurations, 
becoming at most $7\%$ at large evolution times ($t>100$), in contrast to the larger deviations observed in $\Delta_{2,4}(t)$
during evolution. 

\section*{Acknowledgements} 

S.I.M. and P.S. gratefully acknowledge financial support by the Deutsche Forschungsgemeinschaft 
(DFG) in the framework of the
SFB 925 ``Light induced dynamics and control of correlated quantum
systems''. G.M.K and P.S. acknowledge support by the excellence cluster 
`` The Hamburg Center for Ultrafast Imaging: Structure, Dynamics and Control
of Matter at the Atomic Scale'' of the Deutsche Forschungsgemeinschaft. 
P. G. K. gratefully acknowledges the support of  
NSF-PHY-1602994, and the Alexander von Humboldt Foundation.

{}


\begin{thebibliography}{60}
\bibitem{zakharov} V. E. Zakharov, and A. B. Shabat,
Sov. Phys. JETP {\bf 37}, 823 (1973).
 
\bibitem{drummond}  J. F. Corney,  P. D. Drummond,  and A. Liebman,  
Opt. Commun. \textbf{140}, 211 (1997).

\bibitem{yuridavies} Y. S. Kivshar, and D. Luther-Davies,
Phys. Rep. {\bf 298}, 81 (1998). 


\bibitem{pethick} C. J. Pethick and H. Smith,
{\it Bose-Einstein condensation in dilute gases}, Cambridge University
Press (Cambridge, 2002).

\bibitem{stringari} L. P. Pitaevskii, and S. Stringari,
{\it Bose-Einstein Condensation}, Oxford University Press (Oxford, 2003).

\bibitem{djf} D. J. Frantzeskakis,
J. Phys. A Math. Theor. {\bf 43}, 213001 (2010).


\bibitem{siambook} P.~G.~Kevrekidis,
D.~J.~Frantzeskakis, and R.~Carretero-Gonz{\'a}lez,
{\it The Defocusing Nonlinear Schr{\"o}dinger Equation},
SIAM (Philadelphia, 2015). 

\bibitem{chabchoub} A. Chabchoub, O. Kimmoun,
H. Branger, N. Hoffmann, D. Proment, M. Onorato, and N. Akhmediev,
Phys. Rev. Lett. {\bf 110}, 124101 (2013).

\bibitem{colostate} W. Tong, M. Wu, L. D. Carr, and B. A. Kalinikos,
Phys. Rev. Lett. {\bf 104}, 037207 (2010).


\bibitem{burger} S. Burger, K. Bongs, S. Dettmer, W. Ertmer, K. Sengstock, A. Sanpera, G. Shlyapnikov, and M. Lewenstein,
Phys. Rev. Lett. {\bf 83}, 5198 (1999).

\bibitem{Denschlag}  J. Denschlag, J. E. Simsarian, D. L. Feder, C. W. Clark, L. A. Collins, J. Cubizolles, L. Deng, 
E. W. Hagley, K. Helmerson, W. P. Reinhardt, S. L. Rolston, B. I. Schneider, and W. D. Phillips,
Science {\bf 287}, 97 (2000).

\bibitem{becker} C. Becker, S. Stellmer, P. Soltan-Panahi, S. D{\"o}rscher,
M. Baumert, E.-M. Richter, J. Kronj{\"a}ger, K. Bongs, and
K. Sengstock, Nat. Phys. {\bf 4}, 496 (2008).

\bibitem{weller} A. Weller, J. P. Ronzheimer, C. Gross, J. Esteve, M. K. Oberthaler, D. J. Frantzeskakis, G. Theocharis, and 
P. G. Kevrekidis, Phys. Rev. Lett. {\bf 101}, 130401 (2008).

\bibitem{theo} G. Theocharis,
A. Weller, J. P. Ronzheimer, C. Gross, M. K. Oberthaler,
P. G. Kevrekidis, and D. J. Frantzeskakis,
Phys. Rev. A {\bf 81}, 063604 (2010).


\bibitem{Reinhardt}  W. P. Reinhardt, and C. W. Clark, 
J. Phys. B {\bf 30}, L785 (1997).

\bibitem{Scott} T. F. Scott, R. J. Ballagh, and K. Burnett,
J. Phys. B {\bf 31}, L329 (1998).

\bibitem{dutton} Z. Dutton, M. Budde, C. Slowe, and L. V. Hau,
Science \textbf{293}, 663 (2001).

\bibitem{Engels_2007} P. Engels, and C. Atherton,
Phys. Rev. Lett. {\bf 99}, 160405 (2007).

\bibitem{Hakim} Vincent Hakim,
Phys. Rev. E {\bf 55}, 2835 (1997).

\bibitem{Leboeuf} P. Leboeuf, and N. Pavloff,
Phys. Rev. A {\bf 64}, 033602 (2001).

\bibitem{Pavloff} N. Pavloff,
Phys. Rev. A {\bf 66}, 013610 (2002).

\bibitem{Brazhnyi} V. A. Brazhnyi, and A. M. Kamchatnov,
Phys. Rev. A {\bf 68}, 043614 (2003).

\bibitem{Radouani} Abdelaziz Radouani,
Phys. Rev. A {\bf 70}, 013602 (2004).

\bibitem{Carretero} R. Carretero-Gonz{\'a}lez, P. G. Kevrekidis,  D. J. Frantzeskakis, B. A. Malomed, S. Nandi, 
and A. R. Bishop, Math. Comput. Simul. {\bf 74} 361 (2007).

\bibitem{Hans} I. Hans, J. Stockhofe, and P. Schmelcher,
Phys. Rev. A {\bf 92}, 013627 (2015).
 
\bibitem{Syafwan} M. Syafwan, P. Kevrekidis, A. Paris-Mandoki, I. Lesanovsky, P. Kr{\"u}ger, L. Hackermuller, and H. Susanto, 
J. Phys. B: At. Mol. Opt. Phys. {\bf 49}, 235301 (2016).
 
 
\bibitem{Frisch} T. Frisch, Y. Pomeau, and S. Rica,
Phys. Rev. Lett. {\bf 69}, 1644 (1992).

\bibitem{Winiecki} T. Winiecki, J. F. McCann, and C. S. Adams, 
Phys. Rev. Lett. {\bf 82}, 5186 (1999).

\bibitem{Astra} G. E. Astrakharchik, and L. P. Pitaevskii,
Phys. Rev. A {\bf 70}, 013608 (2004).

\bibitem{Landau} L. D. Landau,
J. Phys. (Moscow) {\bf 5}, 71 (1941).


\bibitem{Ketterle_1999} C. Raman, M. K{\"o}hl, R. Onofrio, D. S. Durfee, C. E. Kuklewicz, Z. Hadzibabic, and W. Ketterle,
Phys. Rev. Lett. {\bf 83}, 2502 (1999).

\bibitem{Ketterle_2000} R. Onofrio, C. Raman, J. M. Vogels, J. R. Abo-Shaeer, A. P. Chikkatur, and W. Ketterle,
Phys. Rev. Lett. {\bf 85}, 2228 (2000).


\bibitem{Kamchatnov} A. M. Kamchatnov, and N. Pavloff, 
Phys. Rev. A {\bf 85}, 033603 (2012).

\bibitem{El1} G. A. El, and A. M. Kamchatnov,  
Phys. Lett. A {\bf 350}, 192 (2006).

\bibitem{El2} G. A. El, A. M. Kamchatnov, V. V. Khodorovskii, E. S. Annibale, and A. Gammal,
Phys. Rev. E {\bf 80}, 046317 (2009).

\bibitem{Hoefer} M. A. Hoefer, and B. Ilan,
Phys. Rev. A {\bf 80}, 061601(R) (2009).

\bibitem{hoefer2} G.A. El, M.A. Hoefer,
  Physica D {\bf 333}, 11 (2016).


\bibitem{Huepe} C. Huepe, and M. -E. Brachet, 
C. R. Acad. Sci. Paris {\bf 325}, 195 (1997).


\bibitem{sachadark3} J. Dziarmaga, and K. Sacha,
Phys. Rev. A \textbf{66}, 043620 (2002).

\bibitem{sachadark2} J. Dziarmaga, Z. P. Karkuszewski, and K. Sacha,
Phys. Rev. A \textbf{66}, 043615 (2002).

\bibitem{sachadark1} J. Dziarmaga, Z. P. Karkuszewski, and K. Sacha,
J. Phys. B: At. Mol. Opt. Phys. {\bf 36}, 1217 (2003).

\bibitem{mishmash1}
R. V. Mishmash, and L. D. Carr, Phys. Rev. Lett. {\bf 103}, 140403 (2009). 

\bibitem{mishmash2}  R. V. Mishmash, I. Danshita, C. W. Clark, and L. D. Carr,
  Phys. Rev. A {\bf 80}, 053612 (2009). 

\bibitem{martin} A. D. Martin, and J. Ruostekoski,
Phys. Rev. Lett. {\bf 104}, 194102 (2010).

\bibitem{sacha} D. Delande, and K. Sacha, Phys. Rev. Lett. {\bf 112}, 040402 (2014).

\bibitem{sacha17} A. Syrwid, and K. Sacha, 
Phys. Rev. A {\bf 96}, 043602 (2017). 

\bibitem{Katsimiga} G.  C.  Katsimiga,  G.  M.  Koutentakis,  S.  I.  Mistakidis,  P. G. Kevrekidis,  and  P. Schmelcher,
New  J.  Phys. {\bf 19}, 073004 (2017).

\bibitem{lspp}  S. I. Mistakidis, G. C. Katsimiga, P. G. Kevrekidis, and P. Schmelcher, 
New J. Phys. {\bf 20}, 043052 (2018).  

\bibitem{lgspp} G. C. Katsimiga, G. M. Koutentakis, S. I. Mistakidis, P. G. Kevrekidis, and P. Schmelcher, 
New J. Phys. \textbf{19}, 123012 (2017).  


\bibitem{Wright} K. C. Wright, R. B. Blakestad, C. J. Lobb, W. D. Phillips, and
G. K. Campbell, Phys. Rev. Lett. {\bf 110}, 025302 (2013).

\bibitem{Ryu} C. Ryu, P. W. Blackburn, A. A. Blinova, and M. G. Boshier,  Phys. Rev. Lett. {\bf 111}, 205301 (2013).

\bibitem{Eckel} S. Eckel, J. G. Lee, F. Jendrezejewski, N. Murray, C. W. Clark,
C. J. Lobb, W. D. Phillips, M. Edwards, and C. K. Campbell, Nat. {\bf 506}, 200 (2014).

\bibitem{Jendrezejewski} F. Jendrezejewski, S. Eckel, N. Murray, C. Lanier, M. Edwards,
C. J. Lobb, and C. K. Campbell,  Phys. Rev. Lett. {\bf 113}, 045305 (2014).

\bibitem{Weimer} W. Weimer, K. Morgener, V. P. Singh, J. Siegl, K. Hueck, N. Luick, L. Mathey, and H. Moritz,
Phys. Rev. Lett. {\bf 114}, 095301 (2015).

\bibitem{Roati} A. Burchianti, F. Scazza, A. Amico, G. Valtolina, J. A. Seman, C. Fort, M. Zaccanti, M. Inguscio, and G. Roati,
Phys. Rev. Lett. {\bf 120}, 025302 (2018).


\bibitem{planist} G. Theocharis, P.G. Kevrekidis,
  H.E. Nistazakis, D.J. Frantzeskakis, A.R. Bishop,
  Phys. Lett. A {\bf 337}, 441 (2005).


\bibitem{Fedichev} P. O. Fedichev, and G. V. Shlyapnikov,
Phys. Rev. A {\bf 63}, 045601 (2001).


\bibitem{Mistakidis_per} S.I. Mistakidis, and P. Schmelcher, 
Phys. Rev. A \textbf{95}, 013625 (2017). 


\bibitem{Mistakidis_per1} S.I. Mistakidis, T. Wulf, A. Negretti, and P. Schmelcher, 
J. Phys. B: At. Mol. Opt. Phys. \textbf{48}, 244004 (2015). 



\bibitem{cederbaum1} 
O. E. Alon, A. I. Streltsov, and L. S. Cederbaum, J. Chem. Phys. \textbf{127}, 154103 (2007). 
%
\bibitem{cederbaum2}
O. E. Alon, A. I. Streltsov, and L. S. Cederbaum, Phys. Rev. A \textbf{77}, 033613 (2008).  

\bibitem{khamis} E. G. Khamis, and A. Gammal, Phys. Rev. A {\bf 87}, 045601 (2013).

\bibitem{Kato} M. Kato, X. -F. Zhang, and H. Saito,
Phys. Rev. A {\bf 96}, 033613 (2017).

\bibitem{Pinsker} F. Pinsker, Physica B: Cond. Mat. {\bf 521}, 36 (2017).


\bibitem{Paris} A. Paris-Mandoki, J. Shearring, F. Mancarella, T. M. Fromhold, A. Trombettoni, and P. Kr{\"u}ger,
Sc. Rep. {\bf 7}, 9070 (2017). 


\bibitem{g1d} M. Olshanii, Phys. Rev. Lett. {\bf 81}, 938 (1998).

\bibitem{Inouye} S. Inouye, M. R. Andrews, J. Stenger, H.-J. Miesner, D. M. Stamper-Kurn, and W. Ketterle,
Nat. {\bf 392}, 151 (1998).


\bibitem{Chin} C. Chin, R. Grimm, P. Julienne, and E. Tiesinga,
Rev. Mod. Phys. {\bf 82}, 1225 (2010).

\bibitem{Cao} L. Cao, S. Kr{\"o}nke, O. Vendrell, and P. Schmelcher, 
J. Chem. Phys., {\bf 139}, 134103 (2013).

\bibitem{matakias} S. Kr{\"o}nke, L. Cao, O. Vendrell, and P. Schmelcher, New J. Phys. \textbf{15}, 063018 (2013). 

\bibitem{moulosx}  L. Cao, V. Bolsinger, S. I. Mistakidis, G. M. Koutentakis, S.Kr{\"o}nke, J. M. Schurer, and P. Schmelcher,
J. Chem. Phys. \textbf{147}, 044106 (2017).  

\bibitem{McLachlan} A. D. McLachlan,  Mol. Phys.  \textbf{8}, 39 (1964).

\bibitem{Titulaer} U. M. Titulaer, and R. J. Glauber, Phys. Rev. \textbf{140}, 676 (1965).

\bibitem{Naraschewski} M. Naraschewski, and R. J. Glauber, Phys. Rev. A \textbf{59}, 4595 (1999). 

\bibitem{Sakmann} K. Sakmann, A. I. Streltsov, O. E. Alon, and L. S. Cederbaum,
Phys. Rev. A {\bf 78}, 023615 (2008).

\bibitem{Penrose} O. Penrose, and L. Onsager, Phys. Rev. \textbf{104}, 576 (1956).

\bibitem{Mueller} E. J. Mueller, T. L. Ho, M. Ueda, and G. Baym, Phys. Rev. A \textbf{74}, 33612 (2006). 

\bibitem{Trippenbach} M. Trippenbach, K. G{\'o}ral,  K. Rz{\c{a}}\.{z}ewski, B. A. Malomed, and Y. B. Band, 
J. Phys. B: At. Mol. Opt. Phys. {\bf 33}, 4017 (2000).


\bibitem{Kohn}  W. Kohn, Phys. Rev. {\bf 123}, 1242 (1961).

\bibitem{Albert} M. Albert, T. Paul, N. Pavloff, and P. Leboeuf,
Phys. Rev. Lett. {\bf 100}, 250405 (2008).

\bibitem{comment} Note that in the homogeneous setting the speed of sound, $c$, is uniform. 
Thus, in order to reach region IV
the velocity of the obstacle at the initial stages of the dynamics must be much larger than $c$
resulting to a downstream emission of excitations with velocities comparable to $c$. 

\bibitem{negative} S.I. Mistakidis, L. Cao, and P. Schmelcher, 
Phys. Rev. A \textbf{91}, 033611 (2015). 

\bibitem{spatial} T. Pla{\ss}mann, S.I. Mistakidis, and P. Schmelcher, 
\textbf{arXiv:1802.06693} (2018). 


\bibitem{zorzetos} G.M. Koutentakis, S.I. Mistakidis, and P. Schmelcher, 
\textbf{arXiv:1804.07199} (2018). 


\bibitem{Lode} A. U. Lode, and C. Bruder, 
Phys. Rev. Lett. \textbf{118}, 013603 (2017).  

\bibitem{kaspar} K. Sakmann,  and  M.  Kasevich,  Nat.  Phys. {\bf 12},  451 (2016).

\bibitem{Chatterjee}   B.  Chatterjee,   and  A.  U.  Lode,
{\bf arXiv:1708.07409} (2017). 

\bibitem{Syrwid_shots} A. Syrwid, M. Brewczyk, M. Gajda, and K. Sacha, 
Phys. Rev. A \textbf{94}, 023623 (2016). 

\bibitem{el} G. A. El, A. Gammal, and A. M. Kamchatnov,
Phys. Rev. Lett. {\bf 97}, 180405 (2006).

\bibitem{bettina} B. Gertjerenken, P. G. Kevrekidis, R. Carretero-Gonz{\'a}lez, and B. P. Anderson,
Phys. Rev. A {\bf 93}, 023604 (2016).

\bibitem{kali} E. C. Samson, K. E. Wilson, Z. L. Newman, and B. P. Anderson,
Phys. Rev. A {\bf 93}, 023603 (2016).
  
\bibitem{kaib} N. Meyer, H. Proud, M. Perea-Ortiz, C. O’Neale, M. Baumert,
M. Holynski, J. Kronj{\"a}ger, G. Barontini, and K. Bongs,
Phys. Rev. Lett. {\bf 119}, 150403 (2017).

\bibitem{Bland} T. Bland, K. Paw\l{}owski, M. J. Edmonds,  K. Rz{\c{a}}\.{z}ewski, and N. G. Parker, 
Phys. Rev. A {\bf 95}, 063622 (2017).


\bibitem{Proukakis} B. Jackson, N. P. Proukakis, and C. F. Barenghi,  
Physical Review A, {\bf 75}, 051601 (2007).  

\bibitem{shinn}  G. Moon, W.J. Kwon, H. Lee, Y. Shin,
  Physical Review A {\bf 92}, 051601(R) (2015). 

\bibitem{BF} P. Siegl, S. I. Mistakidis, and P. Schmelcher, 
Phys. Rev. A {\bf 97}, 053626 (2018). 

\end{thebibliography}
\end{document}